%% file: ms.tex
\documentclass[
a4paper,     
12pt,        
article,
onecolumn,   
openany,     
]{memoir}

\usepackage[utf8]{inputenc}

\usepackage[T1]{fontenc}
\usepackage{lmodern}

\usepackage[english]{babel}

\usepackage[final]{microtype}

\usepackage{amsmath,amssymb,mathtools}

\usepackage{pifont}

\usepackage{graphicx}
\usepackage{makeidx}
\usepackage{import}
\usepackage{algorithm}
\usepackage{algpseudocode}

\algdef{SE}[SUBALG]{Indent}{EndIndent}{}{\algorithmicend\ }

\algtext*{Indent}
\algtext*{EndIndent}

\algnewcommand{\LineComment}[1]{\State \(\triangleright\) #1}

\setlrmarginsandblock{0.15\paperwidth}{*}{1}
\setulmarginsandblock{0.2\paperwidth}{*}{1}

\setpnumwidth{3em}
\setrmarg{4em}
\checkandfixthelayout

\newlength\forceindent
\maxsecnumdepth{paragraph}
\setsecnumdepth{paragraph}

\usepackage{titlesec}
\usepackage{etoolbox}

\titlespacing\section{0pt}{12pt plus 4pt minus 2pt}{0pt plus 2pt minus 2pt}
\titlespacing\subsection{0pt}{12pt plus 2pt minus 1pt}{0pt plus 1pt minus 1pt}
\titlespacing\subsubsection{0pt}{12pt plus 2pt minus 1pt}{0pt plus 1pt minus 1pt}

\counterwithout{section}{chapter}
\titleformat*{\section}{\Large\bfseries}

\setsecheadstyle{\normalfont\large\bfseries}
\setsubsecheadstyle{\normalfont\normalsize\bfseries}
\setparaheadstyle{\normalfont\normalsize\bfseries}
\setparaindent{0pt}\setafterparaskip{0pt}

\captiondelim{\space}
\captionnamefont{\small\bfseries}
\captiontitlefont{\small\normalfont}

\changecaptionwidth        
\captionwidth{1\textwidth} 

\usepackage{tabularx}

\makepagestyle{memoirStylePages}
\makerunningwidth{memoirStylePages}{\textwidth}

\makeheadrule{memoirStylePages}{\textwidth}{\normalrulethickness}
\makefootrule{memoirStylePages}{\textwidth}{\normalrulethickness}{0pt}

\makeevenfoot{memoirStylePages}{}{\bfseries\thepage}{}
\makeoddfoot{memoirStylePages}{}{\bfseries\thepage}{}
\makeevenhead{memoirStylePages}{}{\textit{ESCAPE - WP1: Weather \& Climate Dwarfs - D1.1: Batch no. 1 of Dwarfs}}{}
\makeoddhead{memoirStylePages}{}{\textit{ESCAPE - WP1: Weather \& Climate Dwarfs - D1.1: Batch no. 1 of Dwarfs}}{}

\makeatletter
\makepsmarks{memoirStylePages}{
\createmark{chapter}{both}{shownumber}{\@chapapp\ }{ \quad }
\createmark{section}{right}{shownumber}{}{ \quad }
\createplainmark{toc}{both}{\contentsname}
\createplainmark{lof}{both}{\listfigurename}
\createplainmark{lot}{both}{\listtablename}
\createplainmark{bib}{both}{\bibname}
\createplainmark{index}{both}{\indexname}
\createplainmark{glossary}{both}{\glossaryname}
}
\makeatother

\aliaspagestyle{chapter}{chap}

\usepackage{pdfpages}

\usepackage{xspace}
\usepackage{xcolor}
\usepackage{tikz}
\definecolor{gray}{rgb}{0.4,0.4,0.4}
\definecolor{lightgray}{rgb}{0.9,0.9,0.9}
\definecolor{darkblue}{rgb}{0.0,0.0,0.6}
\definecolor{cyan}{rgb}{0.0,0.6,0.6}
\definecolor{maroon}{rgb}{0.5,0.0,0.0}
\definecolor{darkgreen}{rgb}{0.0,0.5,0.0}

\usepackage{listings}
\usepackage{lstautogobble}

\lstdefinestyle{BashStyle}
{
  language=bash,
  basicstyle=\footnotesize\ttfamily,
  frame=single,
  columns=fullflexible,
  backgroundcolor=\color{yellow!10},
  linewidth=\linewidth,
  xleftmargin=0.05\linewidth,
  keepspaces=true,
  framesep=5pt,
  rulecolor=\color{black!30},
  aboveskip=10pt,
  autogobble=true
}

\lstdefinelanguage{XML}
{
  basicstyle=\ttfamily\footnotesize,
  morestring=[b]",
  moredelim=[s][\bfseries\color{maroon}]{<}{\ },
  moredelim=[s][\bfseries\color{maroon}]{</}{>},
  moredelim=[l][\bfseries\color{maroon}]{/>},
  moredelim=[l][\bfseries\color{maroon}]{>},
  morecomment=[s]{<?}{?>},
  morecomment=[s]{<!--}{-->},
  commentstyle=\color{gray},
  stringstyle=\color{orange},
  identifierstyle=\color{darkblue},
  showstringspaces=false
}

\lstdefinestyle{XMLStyle}
{
  language=make,
  basicstyle=\ttfamily\footnotesize,
  numbers=left,
  numberstyle=\tiny,
  numbersep=3pt,
  frame=,
  columns=fullflexible,
  backgroundcolor=\color{black!05},
  linewidth=\linewidth,
  xleftmargin=0.05\linewidth,
  keepspaces=true
}
\lstdefinestyle{CStyle}{
  belowcaptionskip=1\baselineskip,
  breaklines=true,
  frame=single,
  escapeinside={\%*}{*)},
  xleftmargin=\parindent,
  language=C,
  captionpos=b,
  keepspaces=true,
  backgroundcolor=\color{black!05},
  showstringspaces=false,
  numbers=left,
  numbersep=5pt,
  numberstyle=\tiny\color{black},
  basicstyle=\scriptsize\ttfamily,
  keywordstyle=\bfseries\color{green!40!black},
  commentstyle=\itshape\color{purple!40!black},
  identifierstyle=\color{blue},
  stringstyle=\color{orange},
  tabsize=4
}

\lstdefinestyle{CStyleNoLine}{
  belowcaptionskip=1\baselineskip,
  breaklines=true,
  frame=single,
  escapeinside={\%*}{*)},
  xleftmargin=\parindent,
  language=C,
  captionpos=b,
  keepspaces=true,
  backgroundcolor=\color{black!05},
  showstringspaces=false,
  basicstyle=\scriptsize\ttfamily,
  keywordstyle=\bfseries\color{green!40!black},
  commentstyle=\itshape\color{purple!40!black},
  identifierstyle=\color{blue},
  stringstyle=\color{orange},
  tabsize=4
}

\lstdefinestyle{FStyle}{
  belowcaptionskip=1\baselineskip,
  breaklines=true,
  frame=single,
  escapeinside={\%*}{*)},
  xleftmargin=\parindent,
  language=[90]Fortran,
  captionpos=b,
  keepspaces=true,
  backgroundcolor=\color{red!05},
  showstringspaces=false,
  numbers=left,
  numbersep=5pt,
  numberstyle=\tiny\color{black},
  basicstyle=\scriptsize\ttfamily,
  keywordstyle=\bfseries\color{red!40!black},
  commentstyle=\itshape\color{green!40!black},
  identifierstyle=\color{blue},
  stringstyle=\color{orange},
  tabsize=4
}

\lstdefinestyle{FStyleNoLine}{
  belowcaptionskip=1\baselineskip,
  breaklines=true,
  frame=single,
  escapeinside={\%*}{*)},
  xleftmargin=\parindent,
  language=[90]Fortran,
  captionpos=b,
  keepspaces=true,
  backgroundcolor=\color{red!05},
  showstringspaces=false,
  basicstyle=\scriptsize\ttfamily,
  keywordstyle=\bfseries\color{red!40!black},
  commentstyle=\itshape\color{green!40!black},
  identifierstyle=\color{blue},
  stringstyle=\color{orange},
  tabsize=4
}

\ifdefined\HCode

\else

\fi

\ifdefined\HCode
\newcommand{\inlsh}[1]{\texttt{#1}}
\else
\newcommand{\inlsh}[1]{\tikz[anchor=base,baseline]\node[inner sep=2pt,
outer sep=0,draw=yellow!10,fill=yellow!10]{\texttt{#1}};}
\fi

\newcommand{\Atlas}{{\em Atlas}\xspace}

\usepackage{environ}
\usepackage[tikz]{bclogo}
\usetikzlibrary{calc}

\ifdefined\HCode
\NewEnviron{notebox}{\textbf{Note:} \BODY}
\NewEnviron{warningbox}{\textbf{Warning:} \BODY}
\NewEnviron{tipbox}{\textbf{Tip:} \BODY}
\NewEnviron{custombox}[3]{\textbf{#1} \BODY}
\else
\NewEnviron{notebox}
  {\par\medskip\noindent
  \begin{tikzpicture}
    \node[inner sep=5pt,fill=black!10,draw=black!30] (box)
    {\parbox[t]{.99\linewidth}{%
      \begin{minipage}{.1\linewidth}
      \centering\tikz[scale=1]\node[scale=1.5]{\bcinfo};
      \end{minipage}%
      \begin{minipage}{.85\linewidth}
      \textbf{Note}\par\smallskip
      \BODY
      \end{minipage}\hfill}%
    };
   \end{tikzpicture}\par\medskip%
}
\NewEnviron{warningbox}
  {\par\medskip\noindent
  \begin{tikzpicture}
    \node[inner sep=5pt,fill=red!10,draw=black!30] (box)
    {\parbox[t]{.99\linewidth}{%
      \begin{minipage}{.1\linewidth}
      \centering\tikz[scale=1]\node[scale=1.5]{\bcdanger};
      \end{minipage}%
      \begin{minipage}{.85\linewidth}
      \textbf{Warning}\par\smallskip
      \BODY
      \end{minipage}\hfill}%
    };
   \end{tikzpicture}\par\medskip%
}
\NewEnviron{tipbox}
  {\par\medskip\noindent
  \begin{tikzpicture}
    \node[inner sep=5pt,fill=green!10,draw=black!30] (box)
    {\parbox[t]{.99\linewidth}{%
      \begin{minipage}{.1\linewidth}
      \centering\tikz[scale=1]\node[scale=1.5]{\bclampe};
      \end{minipage}%
      \begin{minipage}{.85\linewidth}
      \textbf{Tip}\par\smallskip
      \BODY
      \end{minipage}\hfill}%
    };
   \end{tikzpicture}\par\medskip%
}
\NewEnviron{custombox}[3]
  {\par\medskip\noindent
  \begin{tikzpicture}
    \node[inner sep=5pt,fill=#3!10,draw=black!30] (box)
    {\parbox[t]{.99\linewidth}{%
      \begin{minipage}{.1\linewidth}
      \centering\tikz[scale=1]\node[scale=1.5]{#2};
      \end{minipage}%
      \begin{minipage}{.85\linewidth}
      \textbf{#1}\par\smallskip
      \BODY
      \end{minipage}\hfill}%
    };
   \end{tikzpicture}\par\medskip%
}
\fi

\makeindex

\usepackage[linktoc=all,hyperfootnotes=false,bookmarksdepth=3]{hyperref}
\hypersetup{
colorlinks,
citecolor=darkblue,
filecolor=darkblue,
linkcolor=darkblue,
urlcolor=darkblue,
pdfborder={0 0 0},
pdfauthor={I am the Author}
}
\usepackage{memhfixc}

\ifdefined\HCode
\else
\makeatletter
\newlength\drop

\newcommand*{\titlepage}{
    \thispagestyle{empty}
    \begingroup
    \drop = 0.1\textheight
    \vspace*{\baselineskip}
    \vfill
    \hbox{
      \hspace*{0.1\textwidth}
      \rule{1pt}{\dimexpr\textheight-28pt\relax}
      \hspace*{0.05\textwidth}
      \parbox[b]{0.85\textwidth}{
        \vbox{
          \vspace{\drop}
          {\LARGE\bfseries\raggedright\@title\par}
          \vskip4\baselineskip
          {\HUGE\bfseries \textcolor{darkblue}{ESCAPE}\par}
          \vskip1.0\baselineskip
          {\large\bfseries\@date\par}
          \vspace{0.1\textheight}
          {\small\noindent\@author}\\[\baselineskip]
        }
      }
    }
    \vfill
    \null
\endgroup}
\makeatother
\fi

\begin{document}

\title{Energy-efficient Scalable Algorithms for Weather Prediction at Exascale}
\author{Research and Innovation Action \newline
H2020-FETHPC-2014 \newline
Author: Daniel Thiemert \newline
Date: \today \newline
Project Coordinator: Dr. Peter Bauer (ECMWF) \newline
Project Start Date: 01/10/2015 \newline
Project Duration: 36 months \newline
Published by the ESCAPE Consortium \newline
Version: 0.1 \newline
Contractual Delivery Date: 30/06/2016 \newline
Work Package/ Task: WP1/ T1.1 \newline
Document Owner: ECMWF \newline
Dissemination level: Public \newline
Contributors: Gianmarco Mengaldo, Willem Deconinck, Michail Diamantakis, Alastair McKinstrey, Piet Termonia, Peter Bauer, Nils Wedi }

\frontmatter

\includepdf[page=1]{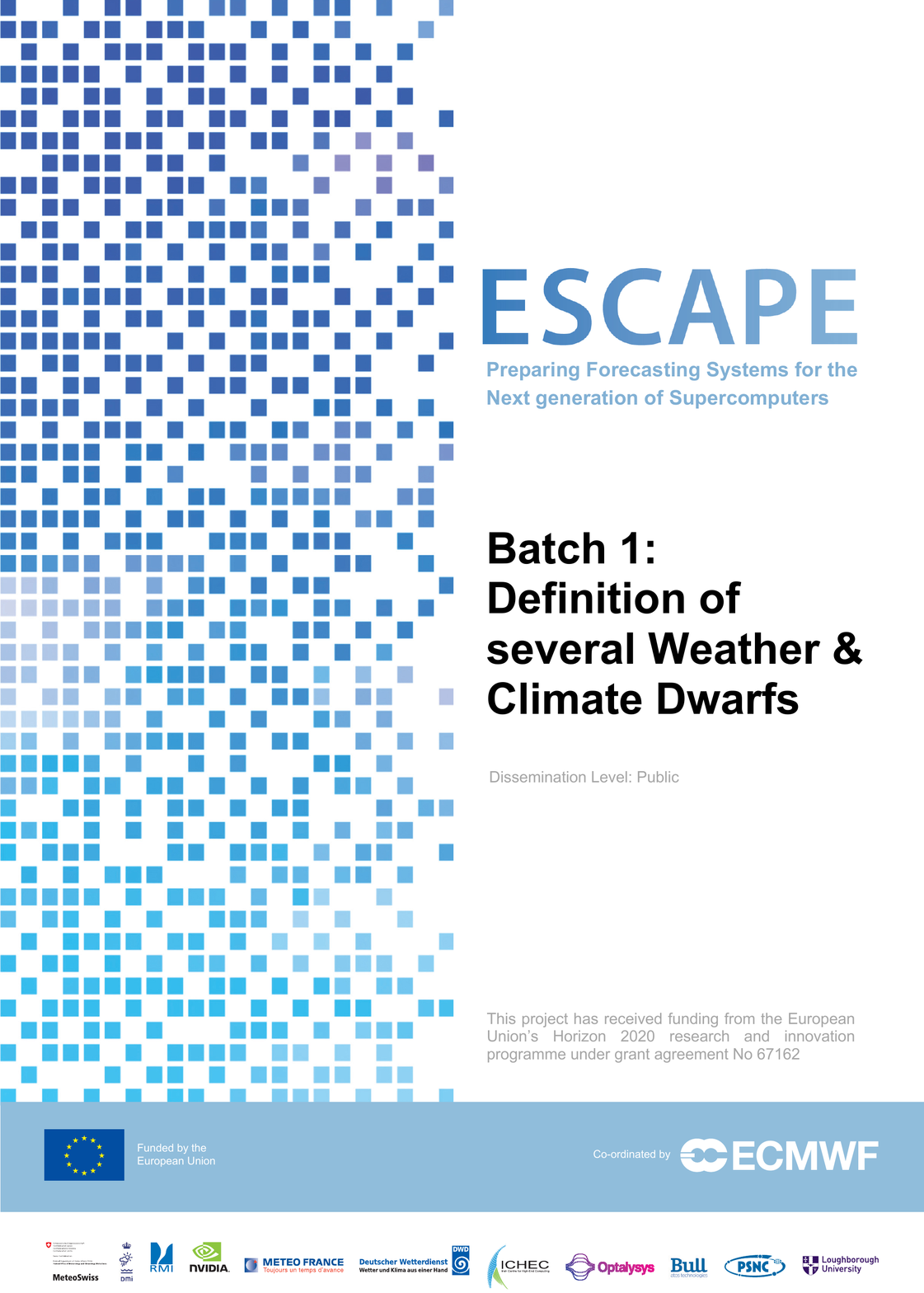}
\includepdf[page=2]{ESCAPE-D1-1-V0-1}
\setcounter{page}{1}


\setcounter{tocdepth}{2}
\ifx\HCode\undefined
\tableofcontents*
\fi

\mainmatter

\section{\label{ES}Executive summary}
This deliverable contains the description of the characteristics of the so-called numerical weather \& climate prediction dwarfs that form key functional components of prediction models in terms of the science that they encapsulate and in terms of computational cost they impose on the forecast production. The ESCAPE work flow between work packages centres on these dwarfs and hence their selection, their performance assessment, code adaptation and optimization is crucial for the success of the project. At this stage of ESCAPE, a selection of established and new dwarfs has been made, their documentation been compiled and the software been made available on the software exchange platform. The selection of dwarfs will be extended throughout the course of the project (see Deliverable D1.2).

The current selection includes the spectral transforms, the cloud microphysics scheme, two and three-dimensional elliptic solvers, a bi-Fourier spectral transform, an interpolation needed for the semi-Lagrangian advection scheme and a first version of the semi-Lagrangian advection scheme itself. This deliverable includes their scientific description and the guidance for installation, execution and testing. This documentation is equivalent to the one available from the ESCAPE Confluence web-page that disseminates the project outcomes.

\section{\label{IN}Introduction}
\subsection{\label{BG}Background}
ESCAPE stands for Energy-efficient Scalable Algorithms for Weather Prediction at Exascale. The project develops world-class, extreme-scale computing capabilities for European operational numerical weather prediction and future climate models. ESCAPE addresses the ETP4HPC Strategic Research Agenda 'Energy and resiliency' priority topic, promoting a holistic understanding of energy-efficiency for extreme-scale applications using heterogeneous architectures, accelerators and special compute units by:
\begin{itemize}
\item Defining and encapsulating the fundamental algorithmic building blocks underlying weather and climate computing;
\item Combining cutting-edge research on algorithm development for use in extreme-scale, high-performance computing applications, minimising time- and cost-to-solution; 
\item Synthesising the complementary skills of leading weather forecasting consortia, university research, high-performance computing centres, and innovative hardware companies.
\end{itemize}
ESCAPE is funded by the European Commission's Horizon 2020 funding framework under the Future and Emerging Technologies - High-Performance Computing call for research and innovation actions issued in 2014.

\subsection{\label{DM}Dwarf map}
A numerical weather prediction (NWP) and/or climate model is composed 
by many interacting building blocks. 
A macro description of such a model can be achieved by separating the 
physical processes that can be fully resolved from those that need to be 
parametrised. This subdivision leads to two macro areas for an NWP and 
Climate model: the \textit{Dynamical core} and the \textit{Physical parametrisation}.

The first is constituted by a set of prognostic partial differential equations (PDEs)
governing the fluid motion, usually the compressible Euler equations (given the 
unfeasibility of solving the compressible Navier-Stokes for the spatial resolutions 
required by operational NWP and Climate applications).

The second encapsulates the subgrid-scale processes that need to be parametrised.
This parametrisation describes the statistical effects of the subgrid-scale processes 
on the mean flow as a function of resolved-scale quantities \cite{malardel2016does}. 

With the continuous increment in spatial resolution \cite{Wedi2013increasing}, some 
additional complexities are being considered as part of an NWP and Climate model, 
such as the land-atmosphere coupling or the ocean-atmosphere coupling, leading 
to a more comprehensive Earth-System model. Given the target of the ESCAPE project, 
the NWP and Climate model considered goes towards the latter direction and the macro 
areas composing the model are the following three:
\begin{itemize}
\item \textit{Dynamical core}, indicated with the letter D;
\item \textit{Physical parametrisation}, denoted with the letter P;
\item \textit{Coupling}, denoted with the letter C.
\end{itemize}
In addition to these three main building blocks we do support some dwarfs that constitute the infrastructure 
of NWP \& climate models. These are denoted by the letter `I' and are intended 
to test different implementations of a heavily used routine within a model, but not 
necessarily constituting a fundamental building block.

\begin{figure}[h]
\centering
\includegraphics[width=0.95\textwidth]{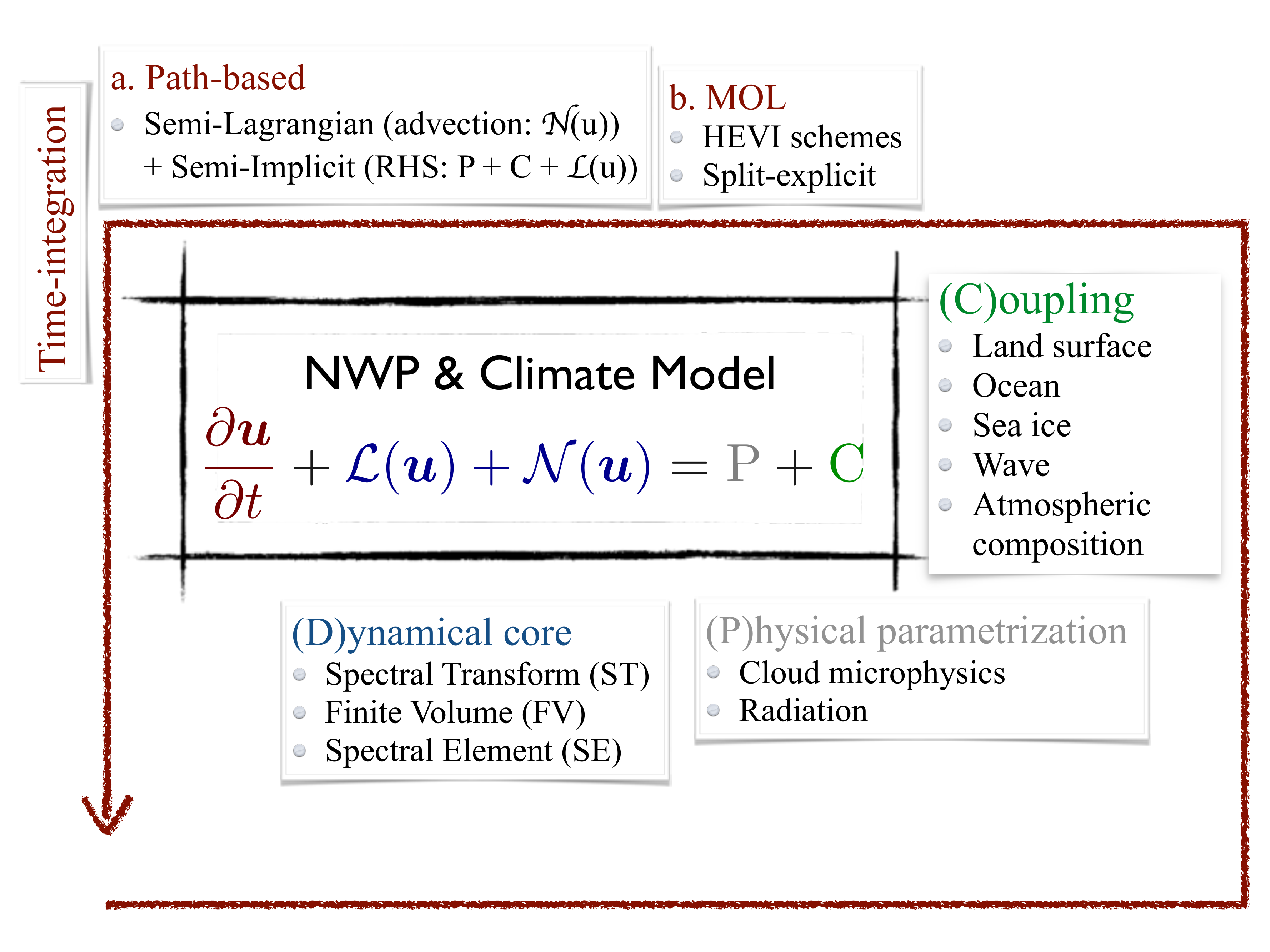}
\caption{Macro (also referred to as top-level) building blocks of an NWP 
and Climate model.}
\label{fig:1}
\end{figure}

Figure~\ref{fig:1} shows a sketch of the three macro building blocks. 
In particular, in blue, we represent the dynamical core that is composed 
by the linear ($\mathcal{L}(\boldsymbol{u})$) and nonlinear ($\mathcal{N}(\boldsymbol{u})$) 
spatial terms of the compressible Euler equations, the latter usually being 
the advection term.
The dynamical core can be either hydrostatic or non-hydrostatic (see also 
\cite{marras2015review}). In the hydrostatic model, the vertical momentum 
is not solved as a prognostic variable in the compressible Euler system 
of PDE equations  but it is rather parametrised and considered as a diagnostic 
variable. 
In the non-hydrostatic model, the vertical momentum is included in the PDE 
system, thus it is considered a prognostic variable. In figure~\ref{fig:1}, we 
include three different spatial discretisations that will be explored as part 
of the ESCAPE project for discretising the linear and nonlinear spatial terms 
of the NWP and Climate equations, namely 
\begin{enumerate}
\item Spectral Transform (ST), that is the current operational system at ECMWF 
and constitutes the numerical technology of many operational models;
\item Finite Volume (FV), that is being considered by a few operational
centres as a potentially competitive candidate for next generation dynamical 
core, given its compact nature and the need for local communication 
only, thus allowing better scalability properties than ST-based models;
\item Spectral Element (SE), that is being considered by few research 
centres and universities, as a potential candidate for next generation 
dynamical core, since it shares the same benefit of FV and allow for 
arbitrary spatial order of accuracy.
\end{enumerate}
To continue the description of figure~\ref{fig:1}, in grey, we denote the physical 
parametrisation. Here, we consider two building blocks, the cloud microphysics 
and the radiation package. They are two essential parts of the atmospheric component of any NWP or climate model. We start in this deliverable 1.1 with the cloud microphysics (Section \ref{sec:cloudMicrophysics}). Radiation will be addressed later in deliverable 1.4 because it is closely related to the interpolation between different meshes that will be developed in that deliverable.

The atmosphere is often coupled to other components such as land surface, ocean, sea ice, wave and atmospheric composition. These components are symbolized in green with the letter C.

\begin{figure}[h]
\centering
\includegraphics[width=0.95\textwidth]{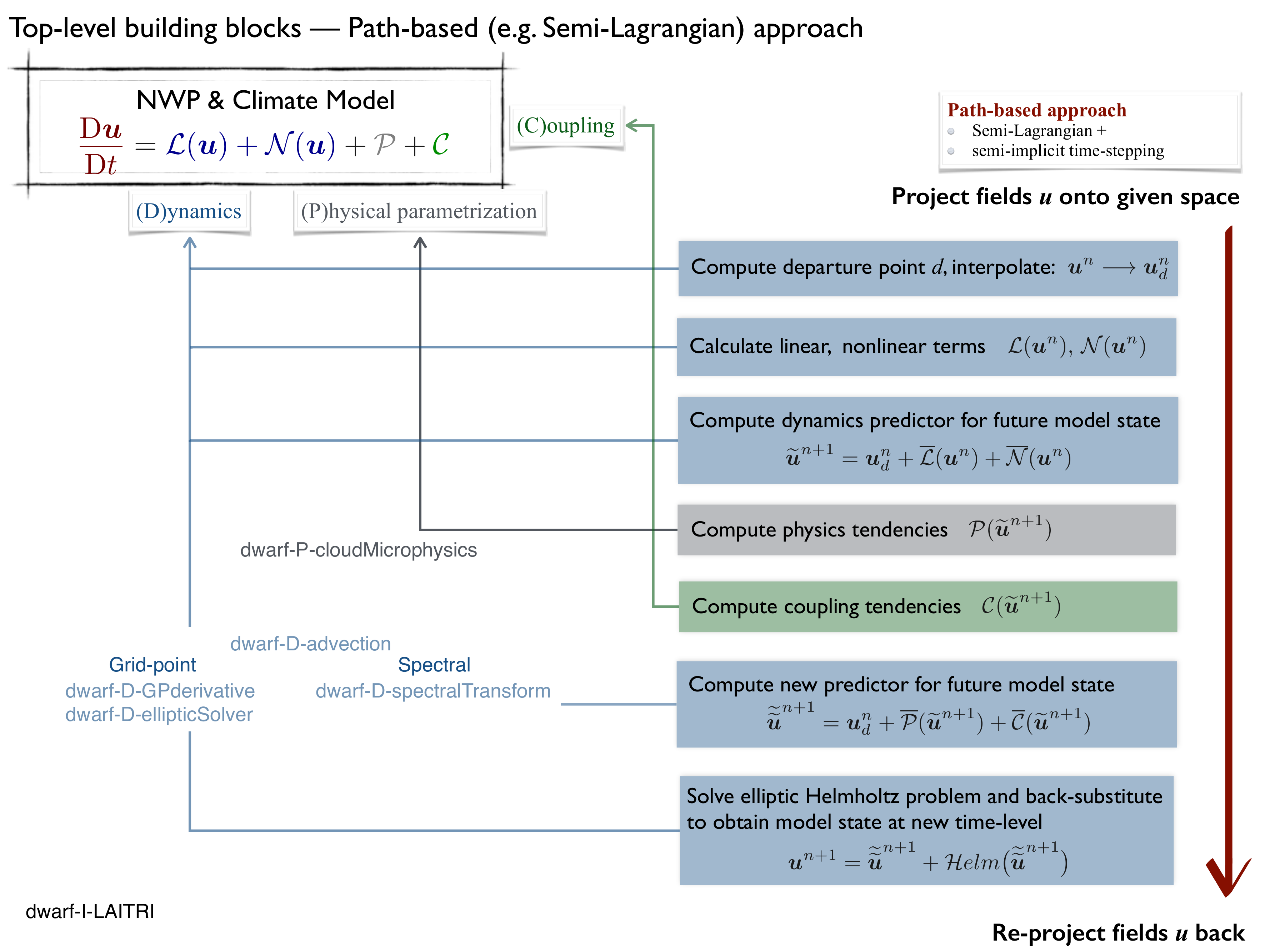}
\caption{Detailed roadmap of the dwarfs in the ESCAPE project.}
\label{fig:2}
\end{figure}

Finally and extremely important, we depict, in red, the time-stepping. This 
is a key aspect for any NWP and Climate model and it needs to be investigated 
carefully. The time-stepping determines which building blocks are required and 
which are not. In particular, in figure~\ref{fig:2}, we show a path-based time-stepping 
approach based on the semi-implicit semi-Lagrangian (SISL) strategy used in the 
current operational system at ECWMF and widely used in many other operational 
centres. 
Path-based time-stepping approaches requires one to calculate the advection 
(or nonlinear) terms separately from the right-hand side terms composed 
by the physical parametrisation, the coupling and the linear terms, that are 
instead calculated in a semi-implicit manner.

In blue, we depict the blocks concerning to the dynamical core and the related
dwarfs. The latter are reported in the following:
\begin{description}
\item \textit{dwarf-D-advection}: within this area of investigation, we currently 
support the implementation of the semi-Lagrangian approach \cite{staniforth1991semi} 
and we will soon deliver the MPDATA (multidimensional positive definite advection 
transport algorithm) \cite{smolarkiewicz1998mpdata} approach; These are called:
\begin{itemize}
\item \textit{dwarf-D-advection-semiLagrangian}
\item \textit{dwarf-D-advection-MPDATA}
\end{itemize}
\item \textit{dwarf-D-spectralTransform}: within this area of investigation, we 
currently support the implementation of the spherical harmonics spectral transform 
strategy adopted in the ECMWF operational system as well as a limited area 
model (LAM) based on the bi-Fourier transform. These are called:
\begin{itemize}
\item \textit{dwarf-D-spectralTransform-sphericalHarmonics}
\item \textit{dwarf-D-spectralTransform-BiFourier}
\end{itemize}
\item \textit{dwarf-D-GPderivative}: within this area of investigation, 
we currently do not support any dwarf. However, as part of another 
European project, Pantarhei, we have implemented second-order 
finite-volume grid-point (GP) derivatives that are used within the 
\textit{dwarf-D-ellipticSolver}.
\item \textit{dwarf-D-ellipticSolver}: within this area of investigation, we 
currently support the implementation of the GCR (generalised conjugate residual) 
approach that is being used within the finite-volume approach developed as 
part of the Pantarhei project. This can be used in conjunction with grid-point 
based spatial discretisations, such as the already mentioned finite-volume 
approach and spectral element methods. The GCR-based dwarf is called:
\begin{itemize}
\item \textit{dwarf-D-ellipticSolver-GCR}
\end{itemize}
\end{description}

In grey, we depict the blocks concerning to the physical parametrisation 
and the related dwarfs. So far we have a dwarf for the cloud microphysics:
\begin{description}
\item \textit{dwarf-P-cloudMicrophysics}: within this area of investigation, 
we currently support the implementation of the scheme used in the ECMWF 
operational system. This is called:
\begin{itemize}
\item \textit{dwarf-P-cloudMicrophysics-IFSScheme}
\end{itemize}
\end{description}

Finally, in green, we depict the blocks concerning to the coupling and the related
dwarfs. Within this area of investigation, we currently 
do not support the implementation of any dwarf. We might add a dwarf at a later stage of this project depending on the progress in the other areas of investigation.

Note the nomenclature adopted throughout. \textit{dwarf-D} denotes a dwarf related 
to the dynamical core, \textit{dwarf-P} denotes a dwarf related to the physical parametrisation 
and \textit{dwarf-C} denotes a dwarf related to the coupling. In addition, after one 
of these three letters, we report the macro area of investigation within the NWP \& 
climate model, that can be dwarf-D-\textit{advection}, dwarf-D-\textit{spectralTransform}, 
dwarf-P-\textit{cloudMicrophysics}. 
These can then explore different strategies to solve the specific macro area of investigation.
Some examples range from the dwarf-D-advection-\textit{semiLagrangian} to the 
dwarf-P-cloudMicrophysics-\textit{IFSScheme}.The last keyword, ultimately identify 
the dwarf. Note that each dwarf is identified by 4 keywords and can have various 
prototype implementations targeting, for instance, different hardware or slightly 
different implementations.

\subsection{\label{SC}Scope of this deliverable}
\subsubsection{\label{OB}Objectives of this deliverable}
This document accompanies the prototype implementations of the first batch of weather and climate dwarfs developed in WP1. The dwarfs initiate the first iteration of the ESCAPE project, feeding into WPs 2, 3, and 4 for code adaptation, hybrid computing and benchmarking and diagnostics. The document aims to provide a documentation of the provided dwarfs to ensure that they are easily usable by the respective partners.
The dwarf implementations are available at:
\url{https://git.ecmwf.int/projects/ESCAPE}.

\subsubsection{\label{WK}Work performed in this deliverable}
As per the task description in the Description of Action for task 1.1, the work performed in this deliverable included the isolation and packaging of canonical NWP algorithms and internal model workflows, including:
\begin{itemize}
\item Spectral transforms scheme (\textit{dwarf-D-spectralTransform-sphericalHarmonics}, see Section \ref{DOC_spectral})
\item Bi-Fourier spectral transform algorithm (\textit{dwarf-D-spectralTransform-BiFourier}, see Section \ref{DOC_bifourier})
\item Semi-Lagrangian interpolation scheme (\textit{dwarf-D-interpolation-semiLagrangian}, see Section \ref{DOC_slinterpol})
\item Cloud microphysics scheme (\textit{dwarf-P-cloudMicrophysics}, see Section \ref{DOC_cloud})
\end{itemize}
Additionally, newly developed dwarfs have been included, namely:
\begin{itemize}
\item Elliptic solvers (\textit{dwarf-D-ellipticSolver-GCR}, see Section \ref{DOC_elliptic})
\item Semi-Lagrangian advection scheme (\textit{dwarf-D-advection-semiLagrangian}, see Section \ref{DOC_slnew})
\end{itemize}

\subsubsection{\label{DV}Deviations and counter measures}
Deviations and counter measures were not required for the completion of this deliverable.

\section{\label{DEF}Definition of a Dwarf}
A numerical weather \& climate prediction dwarf is a self-contained set of algorithms representing a key functional block of a forecast model. More specifically, a weather \& climate prediction dwarf should encapsulate a relevant characteristic or required functionality of a weather and climate prediction model and it is meant to be a runnable and verifiable mini-application. The specific implementation of a dwarf in ESCAPE is referred to as prototype, where the prototype should represent a new concept or implementation of a weather and climate prediction model. Examples from this perspective are the specialization of a particular dwarf to different hardware (e.g. accelerators) or a new concept allowing the solution of a required functional block in a different manner (e.g. different advection schemes for the dynamical core).

Within the ESCAPE project, the main interest is focussed on energy-efficient strategies targeting next generation high-performance computing (HPC) facilities. Therefore, a dwarf within this context should reflect the following characteristics:
\begin{enumerate}
\item DEFINITION: be a key functional block (i.e. an essential part) of a weather and climate prediction model. Note that also new modelling strategies to deal with required parts of a prediction model are welcome and should be considered within this context (e.g. investigation of different numerical strategies such as semi-Lagrangian and finite-volume schemes).
\item PARALLELISM AND FLEXIBILITY: be intrinsically parallel (unless explicitly stated otherwise) and portable on different machines having different architectures (CPU/GPU/Accelerators). It should also allow for consistent modifications in order to allow partners to test as many solutions as required.
\item VERIFICATION: accurately provide the results expected by the particular key functional block implemented against one or more benchmarks. This point is essential for ensuring the correct behaviour of the various solution strategies and hardware implementations tested.
\item REPRODUCIBILITY: be reproducible across different hardware and different algorithmic strategies up to a predefined tolerance.
\item READIBILITY: be 'easily' accessible to people working in different fields with different expertise. Have a clear documentation that should include,
\begin{enumerate}
\item Definition;
\item Motivation and importance;
\item Objectives of the dwarf;
\item Description of the results one should expect.
\end{enumerate}
\item PROFILING: provide a clear and well-described output in terms of:
\begin{enumerate}
\item relative speed to solution;
\item relative energy to solution;
\item relative cost to solution;
\item 'relative' means that the various algorithmic/hardware solutions must be comparable. 
\end{enumerate}
\item INTEGRATION (optional): describe the interface(s) for possible integration with the other dwarfs to build a complete prediction model. This point, although optional, might be beneficial towards the end of the project, when new solutions strategies could be available.
\end{enumerate}

\section{Description of individual Dwarfs}
\subsection{\label{DOC_spectral}Spectral transform scheme}

\input{dwarf1}

\subsection{\label{DOC_elliptic}Elliptic solvers}

\input{dwarf2}

\subsection{\label{DOC_cloud}Cloud microphysics scheme}
\label{sec:cloudMicrophysics}

\input{dwarf3}

\subsection{\label{DOC_bifourier}Bi-Fourier spectral transform algorithm}

\input{dwarf4}
\subsection{\label{DOC_slinterpol}Semi-Lagrangian interpolation scheme}

\input{dwarf5}

\subsection{\label{DOC_slnew}Semi-Lagrangian advection scheme}

\input{dwarf6}


\section{\label{CON}Conclusions}
A first representative set of weather and climate prediction model sub-components (dwarfs) has been established, documented and made available for uptake and testing by the ESCAPE project partners. The range of computational characteristics of the dwarfs covers various impediments to scalability and efficiency, namely memory bandwidth, communication and computational cost. The completion of this deliverable allows to progress with subsequent tasks of ESCAPE dealing with code adaptation and performance evaluation on the available hardware architectures. The deliverable has been produced on time and its outcome has been disseminated to project partners through the proposed mechanisms of ESCAPE.

\backmatter


\bibliographystyle{plain}
\bibliography{ESCAPE_D1.1_v0}

\cleardoublepage\pagestyle{empty}
\includepdf[page=6]{ESCAPE-D1-1-V0-1}
\includepdf[page=7]{ESCAPE-D1-1-V0-1}

\end{document}

%% file: dwarf1.tex
\subsubsection{Scope}
The spectral transform method based on \textit{spherical 
harmonics} (on the sphere) constitutes one of the core 
building blocks of many global weather \& climate models; 
for instance, the Integrated Forecast System (IFS) at ECMWF 
started using it approximately thirty years ago and it is still the 
underlying algorithm of global operational forecasts run daily 
at ECMWF. The code is also shared with the operational model 
ARPEGE at M\'et\'eo-France and, in parts, with regional models 
HIRLAM and ALADIN.
However, with the increase in resolution, spectral transform 
models were believed to become prohibitively expensive because 
of the relative increase in computational cost of the Legendre 
transforms compared to gridpoint computations. Nevertheless 
the spectral transform method used in IFS continues to run 
efficiently due to the application of (a) highly efficient 
matrix-matrix multiply (DGEMM) operations and (b) the fast 
Legendre transform \cite{Wedi2013}. However, the communication 
costs related to the data transpositions between grid-point 
and spectral space at every time-step remain an issue.

The overarching scope of this dwarf is to understand if this
core building block can be efficiently ported onto next generation 
exascale HPC systems, in an heterogeneous computing environment, 
where the need for extreme parallelism may limit the use of the 
spectral transform method.

As the computational kernels of the spectral transform dwarf 
are just Fourier transforms and matrix-matrix multiplies, we 
would hope to explore how this dwarf could be run at the speed 
of light on an Optalysis optical device.
If successful this would be revolutionary for NWP and keep spectral 
models at the leading edge of scientific research.

\subsubsection{Objectives}
The main objective of this dwarf is to port the underlying code 
to an accelerator or many-core architecture (these will be also 
referred to as {\bfseries devices}) environment and try to match 
or exceed the performance that can today be achieved on conventional 
multi-core (also referred to as {\bfseries host}) systems such as 
the CRAY XC-30 at ECMWF. In particular, it is important to achieve 
a time-to-solution comparable or better than the current implementation 
on host systems while saving energy due to the use of accelerator devices.

The detailed goals are therefore:
\begin{itemize}
\item to measure the time-to-solution provided by implementations 
of the spectral transform on different hardware, 
\item to measure the energy-to-solution and 
\item to find the best compromise that minimises both the time-to-solution 
and the energy-to-solution.
\end{itemize}
To achieve these goals we currently provide two prototype 
implementations; the first, \inlsh{prototype1}, relies on
the Atlas data-structure, while the second, \inlsh{prototype2} 
is directly extracted from the IFS.
These two prototypes currently support standard host architectures 
and multi-threading is achieved through the use of OpenMP.
The development of at least other two prototypes is essential 
to test device-type architectures and the Optalysis optical 
processors.
Note that \inlsh{prototype2} can be used as a benchmark 
for measuring the performance in terms of time-to-solution,
scalability and energy-to-solution as it represents the current 
operational implementation at ECMWF.

\subsubsection{Definition of the Dwarf}
\begin{figure}[htb!]
\centering
\includegraphics[width=0.65\textwidth]{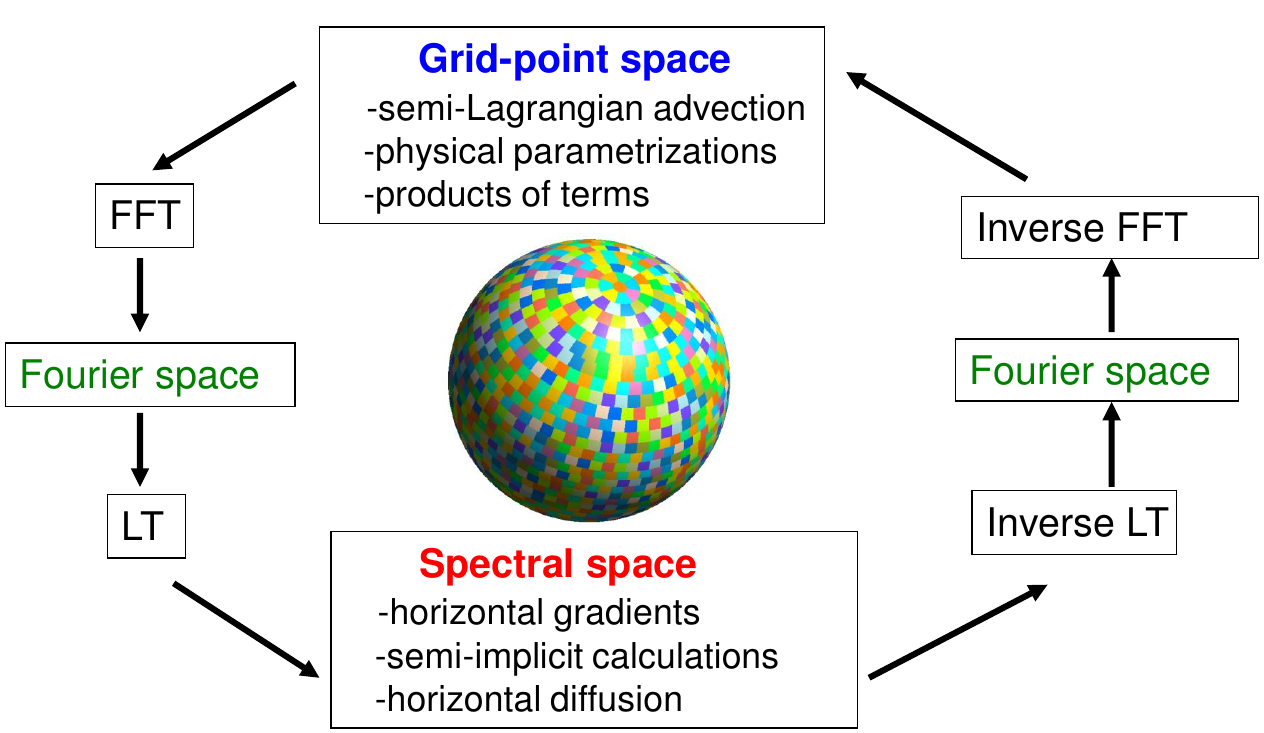}
\caption{Schematics of dwarf-D-spectralTransform-sphericalHarmonics.}
\label{doc_dwarf1:fig:schematics}
\end{figure}

Dwarf-D-spectralTransform-sphericalHarmonics implements the spectral 
transform method on the sphere. This method involves discrete spherical 
harmonics transformations between physical (gridpoint) space and spectral 
(spherical harmonics) space. A sketch representing this dwarf is depicted 
in figure \ref{doc_dwarf1:fig:schematics}.
In particular, following closely the work in \cite{Wedi2013}, a spherical 
harmonics transform is a Fourier transformation in longitude and a Legendre 
transformation in latitude, thus keeping a latitude-longitude structure in gridpoint 
space. The Fourier-transform part is computed numerically by using the fast 
Fourier transform (FFT) that reduces the computational complexity to $\approx 
\mathcal{O}(N^2 \log N)$, where $N$ is the cut-off spectral truncation wavenumber. 
On the other hand, the Legendre transform has a computational complexity 
of $\mathcal{O}(N^3)$ and with increasing horizontal resolution this will become 
the most expensive part of the computation. In the next subsection, we will 
very briefly introduce the fast spherical algorithm introduced in \cite{Wedi2013}, 
that reduces the computational complexity down to $\mathcal{O}(N^2 \log N)$.

\paragraph{The spherical harmonics transform}
If we consider a scalar variable $\psi$ on a vertical level $\ell$, the truncated series 
expansion in spherical harmonics assumes the following form:
\begin{equation}
\psi(\theta, \lambda) = \sum_{m = -M}^{M} e^{i\,m\,\lambda} 
\sum_{n = |m|}^{N(m)} \psi^{m}_{n,\ell} \bar{P}_{n}^{m}[\cos(\theta)],
\label{doc_dwarf1:eq:harmonics-1}
\end{equation}
where $\theta$ and $\lambda$ denote co-latitude ($\phi = 90 - \theta$ is the geographical 
latitude with 0 at the equator) and longitude, respectively; $\psi_{n,\ell}^{m} $are the 
spectral coefficients of the field $\psi$ at level $\ell$; and $\bar{P}_{n}^{m}$ are the 
normalized Legendre polynomials of degree $n$ and order $m$ as a function of 
the latitude only. The indices $M$ and $N(m)$ denote the cut-off spectral truncation 
wavenumber in the spherical harmonics expansion and the choice of N(m) specifies 
the truncation type.

Equation \ref{doc_dwarf1:eq:harmonics-1} represents the discrete inverse spherical harmonics 
transform in spectral coefficient space.

A direct spherical harmonics representation is accomplished by a Fourier transformation 
in longitude as:
\begin{equation}
\psi_{m,\ell}(\theta) = \frac{1}{2\pi} 
\int_{0}^{2\pi}\psi(\lambda, \theta)\,e^{-i\,m\,\lambda}\,d\lambda
\label{doc_dwarf1:eq:harmonics-2}
\end{equation}
and a Legendre transformation in latitude for each $m$ as 
\begin{equation}
\psi^{m}_{n,\ell}(\theta) = \frac{1}{2\pi} 
\int_{0}^{2\pi}\psi_{m,\ell} \bar{P}^{m}_{n}[\cos(\theta)]\, d\cos(\theta)
\label{doc_dwarf1:eq:harmonics-3}
\end{equation}
The Fourier transform is computed using the FFT algorithm detailed in \cite{Temperton1983}. 
The Legendre transforms require the accurate discrete computation of the integral 
in equation \ref{doc_dwarf1:eq:harmonics-3}, that is accomplished by Gaussian quadrature:
\begin{equation}
\psi_{n,\ell}^{m} = \sum_{k=1}^{K} w_{k}\psi_{m,\ell}(x_{k})\bar{P}_{n}^{m}(x_{k})
\label{doc_dwarf1:eq:harmonics-4}
\end{equation}
at the $K = (2N + 1) / 2$ (for a linear grid) special quadrature points (`Gaussian latitudes') 
given by the roots of the ordinary Legendre polynomials $P^{m=0}_{N}(x) = 0$ and the 
Gaussian weights computed from the following equation:
\begin{equation}
w_{k} = \frac{2N + 1}{[\bar{P}^{m=1}_{N}(x_{k}]^{2}}
\label{doc_dwarf1:eq:weights}
\end{equation}
For additional information and detail, the interested reader can refer to \cite{Wedi2013}.

\newpage
\paragraph{Pseudo-algorithm}
In this subsection, we depict the pseudo-algorithm underlying the dwarf.

\begin{algorithm}[H]
\caption{Spectral-transform spherical-harmonics pseudo-algorithm}
\label{doc_dwarf1:alg:sphericalHarmonics}
\begin{algorithmic}
\vspace{.2cm}
\State For any grid-point field $\psi(\theta,\lambda) = \sum_{m = -M}^{M} e^{i\,m\,\lambda} 
\sum_{n = |m|}^{N(m)} \psi^{m}_{n,\ell} \bar{P}_{n}^{m}[\cos(\theta)]$:\\
\For{$m$ = $-M$,...,$M$}
   \vspace{.4cm}
   \State $\displaystyle 
   \psi_{m,\ell}(\theta) = \frac{1}{2\pi} 
   \int_{0}^{2\pi}\psi(\lambda, \theta)\,e^{-i\,m\,\lambda}\,d\lambda$ \hspace{1cm} [FFT]\\
 \For{$n$ = $|m|$,...,$N(m)$}
   \vspace{.4cm}
   \State $\displaystyle 
   \psi^{m}_{n,\ell}(\theta) = \frac{1}{2\pi} 
   \int_{0}^{2\pi}\psi_{m,\ell} \bar{P}^{m}_{n}[\cos(\theta)]\, d\cos(\theta)$ \hspace{1cm} [FLT]\\
 \EndFor
\EndFor
\end{algorithmic}
\end{algorithm}

\paragraph{I/O interfaces }
This dwarf has a simple input/output (I/O) layout. 
Specifically, the interfaces, with reference to algorithm~\ref{doc_dwarf1:alg:sphericalHarmonics} 
in terms of I/O data are as follows:
\begin{itemize}
\item \textbf{Input}: an Atlas-type field in grid-point space;
\item \textbf{Output}: an Atlas-type field in spectral space; 
\end{itemize}
The dimensions of these Atlas-type fields  are determined 
by the grid employed that can be specified as a command-line 
argument. Grids that are reasonable for current operations 
and for next generation global NWP are \textbf{TCo1279} 
(9 km global resolution -- currently operational), \textbf{TCo1999} 
(5 km global resolution), \textbf{TCo3999} (2.5 km global resolution), 
\textbf{TCo7999} (1.3 km global resolution). The acronym \textbf{TCo}
refers to the Cubic Octahedral grid.

To use these grids and on how to run the dwarf, refer to section~\ref{doc_dwarf1:sec:run}.

\subsubsection{Prototypes}
\label{doc_dwarf1:sec:prototypes}
In this section we provide a brief description of each prototype 
implemented.

\paragraph{Prototype 1}
The first prototype, \inlsh{prototype1}, implements the spherical 
harmonics core routines relying on the Atlas data-structure.
The dwarf is divided into two main files:
\begin{itemize}
\item \inlsh{dwarf-D-spectralTransform-sphericalHarmonics-prototype1.F90}, 
which implements the main program, and
\item \inlsh{dwarf-D-spectralTransform-sphericalHarmonics-helper-module.F90},
which contains some support functionalities.
\end{itemize}
The key part of the code is contained within the file \\
\inlsh{dwarf-D-spectralTransform-sphericalHarmonics-prototype1.F90},
inside the subroutine \inlsh{run}, where the calls for the inverse 
and direct transforms are made and repeated several times in a 
\inlsh{do} loop in order to mimic time-stepping.

\paragraph{Prototype 2}
The second prototype, \inlsh{prototype2}, implements the spherical 
harmonics core routines NOT relying on the Atlas data-structure.
The dwarf is encapsulated into one single file \\
\inlsh{dwarf-D-spectralTransform-sphericalHarmonics-prototype2.F90} 
and the main functions and subroutines are directly extracted from 
the IFS. Note that this prototype constitutes a benchmark and we provide 
limited support.

\subsubsection{Dwarf installation and testing}
\label{doc_dwarf1:sec:installation}
In this section we describe how to download and install 
the dwarf along with all its dependencies and we show 
how to run it for a simple test case.

\paragraph{Download and installation}
The first step is to download and install the dwarf along 
with all its dependencies. With this purpose, it is possible 
to use the script provided under the ESCAPE software collaboration 
platform:\\
\url{https://git.ecmwf.int/projects/ESCAPE}.

Here you can find a repository called \inlsh{escape}.
You need to download it. There are two options to do this. One option is to use ssh. For this option you need to add an ssh key to your bitbucket account at \url{https://git.ecmwf.int/plugins/servlet/ssh/account/keys}. The link "SSH keys" on this website gives you instructions on how to generate the ssh key and add them to your account. Once this is done you should first create a 
folder named, for instance, ESCAPE, enter into it 
and subsequently download the repository by using the following the steps below:
\begin{lstlisting}[style=BashStyle]
mkdir ESCAPE
cd ESCAPE/
git clone ssh://git@git.ecmwf.int/escape/escape.git
\end{lstlisting}
The other option to download the repo is by using https instead of ssh. Instead of the git command above you then need to use 
\begin{lstlisting}[style=BashStyle]
git clone https://<username>@git.ecmwf.int/scm/escape/escape.git
\end{lstlisting}
where <username> needs to be replace by your bitbucket username.

Once the repository is downloaded into the \inlsh{ESCAPE} folder 
just created, you should find a new folder called \inlsh{escape}. 
The folder contains a sub-folder called \inlsh{bin} that has the 
python/bash script (called \inlsh{escape}) that needs to be 
run for downloading and installing the dwarf and its dependencies. 
To see the various options provided by the script you can type:
\begin{lstlisting}[style=BashStyle]
./escape/bin/escape -h
\end{lstlisting}
To download the dwarf you need to run 
the following command:
\begin{lstlisting}[style=BashStyle]
./escape/bin/escape checkout dwarf-D-spectralTransform-sphericalHarmonics \ 
--ssh
\end{lstlisting}
To use https you need to replace --ssh with --user <username>. The commands above automatically check out the \inlsh{develop}
version of the dwarf. If you want to download a specific branch 
of this dwarf, you can do so by typing:
\begin{lstlisting}[style=BashStyle]
./escape/bin/escape checkout dwarf-D-spectralTransform-sphericalHarmonics \ 
--ssh --version <branch-name>
\end{lstlisting}
An analogous approach can be used for the \inlsh{-\,-user} 
version of the command. You should now have a folder called 
\inlsh{dwarf-D-spectralTransform-sphericalHarmonics}.

In the above command, you can specify several other optional 
parameters. To see all these options and how to use them you 
can type the following command:
\begin{lstlisting}[style=BashStyle]
./escape checkout -h
\end{lstlisting}

At this stage it is possible to install the dwarf 
and all its dependencies. This can be done in two 
different ways. The first way is to compile and 
install each dependency and the dwarf separately:
\begin{lstlisting}[style=BashStyle]
./escape/bin/escape generate-install \ 
dwarf-D-spectralTransform-sphericalHarmonics
\end{lstlisting}
The command above will generate a script called \\ 
\inlsh{install-dwarf-D-spectralTransform-sphericalHarmonics} 
that can be run by typing:
\begin{lstlisting}[style=BashStyle]
./install-dwarf-D-spectralTransform-sphericalHarmonics
\end{lstlisting}
This last step will build and install the dwarf 
along with all its dependencies in the following 
paths:
\begin{lstlisting}[style=BashStyle]
dwarf-D-spectralTransform-sphericalHarmonics/builds/
dwarf-D-spectralTransform-sphericalHarmonics/install/
\end{lstlisting}

The second way is to create a bundle that compiles 
and installs all the dependencies together:
\begin{lstlisting}[style=BashStyle]
./escape/bin/escape generate-bundle \ 
dwarf-D-spectralTransform-sphericalHarmonics
\end{lstlisting}
This command will create an infrastructure to avoid
compiling the single third-party libraries individually
when some modifications are applied locally to one of 
them. After having run the above command for the bundle, 
simply follow the instructions on the terminal to complete 
the compilation and installation process.

In the commands above that generate the installation 
file, you can specify several other optional parameters. 
To see all these options and how to use them you 
can type the following command:
\begin{lstlisting}[style=BashStyle]
./escape/bin/escape generate-install -h
./escape/bin/escape generate-bundle -h
\end{lstlisting}

\paragraph{Testing}
You should now verify that the dwarf works as expected.
With this purpose, we created a testing framework that
allows us to verify that the main features of the dwarf 
are working correctly.

To run this verification, you should run the following 
command:
\begin{lstlisting}[style=BashStyle]
ctest -j<number-of-tasks>
\end{lstlisting}
from inside the \inlsh{builds/dwarf-D-spectralTransform-sphericalHarmonics}
folder.
\begin{warningbox}
We strongly advise you to verify via ctest that 
the main functionalities of the dwarf are working 
properly any time you apply modifications to the 
code. Updates that do not pass the tests cannot 
be merged. 
In addition, if you add a new feature to the dwarf,
this should be supported by a test if the existing
testing framework is not already able to verify its
functionality.
\end{warningbox}
For instructions on how to run the executables 
see the next section.

\subsubsection{Run the dwarf}
\label{doc_dwarf1:sec:run}
In this section we explain how to run the dwarf.
We make a distinction on how to run \inlsh{prototype1} and \inlsh{prototype2}.
Note also that \inlsh{prototype2} should just be used for benchmarking 
purposes.

\paragraph{Prototype1}
To run the first prototype, which is based on the Atlas data-structure, 
you should follow the following steps. Create a folder called \inlsh{test}
where you want to run the dwarf and enter into it.
You should then export the path of the following renamed binary files 
\begin{lstlisting}[style=BashStyle]
testdata=sources/dwarf-D-spectralTransform-sphericalHarmonics-testdata
\end{lstlisting}
as follows:
\begin{lstlisting}[style=BashStyle]
export DWARF_D_SPECTRALTRANSFORM_SPHERICALHARMONICS_TESTDATA_PATH=testdata
\end{lstlisting}
You then need to copy the \inlsh{.json} files in your current directory 
\begin{lstlisting}[style=BashStyle]
cp sources/dwarf-D-spectralTransform-sphericalHarmonics/config-files/*.json .
\end{lstlisting}
Then, with reference to the renamed folder 
\begin{lstlisting}[style=BashStyle]
inst=dwarf-D-spectralTransform-sphericalHarmonics/install/
\end{lstlisting}
you can run the dwarf as follows:
\begin{lstlisting}[style=BashStyle]
inst/bin/dwarf-D-spectralTransform-sphericalHarmonics-prototype1 \ 
--config TL159.json
\end{lstlisting}
where we used as an example the file for the grid TL159.
Note that the test cases provided are the following:
\begin{itemize}
\item {\bfseries TL159} (125 km)
\item {\bfseries TL1279} (16 km)
\item {\bfseries TCo639} (16 km)
\item {\bfseries TCo1279} (9 km), same as the HRES operational model at ECMWF
\item {\bfseries TCo1999} (5 km)
\item {\bfseries TCo3999} (2.5 km)
\item {\bfseries TL7999} (1.3 km)
\end{itemize}
The dwarf test cases are setup to have minimal data input requirements, 
with input fields being replicated for the high resolution cases.

\paragraph{Prototype2}
Batch jobs for the ECMWF's XC-30 cluster (Ivybridge cores) 
are provided in the following folder 
\begin{lstlisting}[style=BashStyle]
testdata=sources/dwarf-D-spectralTransform-sphericalHarmonics-testdata
\end{lstlisting}
The defaults are set so that no additional namelist settings 
are required. However you can make some changes,
\begin{itemize}
\item {\bfseries IMAX\_FLDS\_IN} = <n> the number of fields 
to be read from input files
\item {\bfseries IMAXFLD} = <m> the number of fields to transform. 
For fields m > n the additional fields are copied in a modulo 
fashion from the original
\item {\bfseries ITERS} = <n> changes the number of iterations 
from the default of 100
\item {\bfseries LNORMS} = true (default is false) computes 
spectral norms on each iteration and prints the error relative 
to the initial state
\begin{lstlisting}[style=BashStyle]
        TRANSFORM_TEST initialisation, on    16 tasks, took     1.48 sec
        time step      1 took    1.78 max err  0.944E-14
        time step      2 took    0.03 max err  0.189E-13
        time step      3 took    0.03 max err  0.283E-13
\end{lstlisting}
\item {\bfseries LUSEFLT} = [true,false] controls use of the Fast Legendre 
Transform, which is only beneficial for resolutions TCo1279 and beyond
\end{itemize}

The directory ECMWF\_OUT contains outputs from running the 
jobs on one of the ECMWF XC-30 clusters CCB. The files are
\begin{lstlisting}[style=BashStyle]
-rw-r----- 1 mpm rd  98376 Jul 14 10:03 T159trans.out.FLT=T
-rw-r----- 1 mpm rd  94971 Jul 14 11:03 T159trans.out.FLT=F
\end{lstlisting}
Outputs are provided showing use (or not) of the Fast Legendre 
Transform (FLT), though in practice the FLT is only expected 
to be beneficial at resolutions beyond TCO1279, although 
it is perfectly acceptable to use the FLT at lower resolutions 
for testing purposes by setting the the namelist option LUSEFLT 
to TRUE.
Within these outputs you will find lines like the following 
printed at the end of the job showing the maximum error and 
time for executing 100 iterations of the Spectral Transform
Dwarf.
\begin{lstlisting}[style=BashStyle]
TRANSFORM_TEST timestep loop, on     4 tasks, took     9.69 sec
MAXIMUM ERROR= 0.800E-05
\end{lstlisting}

Ideally the maximum error should be no greater than one order 
of magnitude of the values shown below.\\

With fast Legendre transform disabled (LUSEFLT=false),
\begin{lstlisting}[style=BashStyle]
TL159,   MAXIMUM ERROR= 0.80E-05
TCo639,  MAXIMUM ERROR= 0.16E-11
TCo1279, MAXIMUM ERROR= 0.18E-11
TCo1999, MAXIMUM ERROR= 0.12E-11
TCo3999, MAXIMUM ERROR= 0.16E-10
TL7999,  MAXIMUM ERROR= 0.17E-07
\end{lstlisting}

With fast Legendre transform enabled (LUSEFLT=true),
\begin{lstlisting}[style=BashStyle]
TL159,   MAXIMUM ERROR= 0.80E-05
TCo639,  MAXIMUM ERROR= 0.16E-11
TCo1279, MAXIMUM ERROR= 0.22E-08
TCo1999, MAXIMUM ERROR= 0.17E-09
TCo3999, MAXIMUM ERROR= 0.19E-10
TL7999,  MAXIMUM ERROR= 0.18E-07
\end{lstlisting}
Note that \inlsh{prototype2} is meant for benchmarking purposes, 
therefore we do provide limited documentation and support for it.

\subsubsection{Integration}
Given that the host version of the dwarf is derived from 
the IFS trans library (code cycle CY41R1) there should be a minimal effort 
to integrate any modifications into an official IFS source cycle.

On the other hand, modifications for a GPU device implementation of the dwarf 
are only meaningful if the modifications are consistent with a more complete 
IFS port to a GPU device. Programming productivity should be considered for 
such a port taking into account features such as deep copy of derived type 
objects that are used for example in the fast Lengendre transform implementation. 
The dwarf should always be run with multiple tasks and over multiple nodes, 
to assess the effect of any potential overlap of computations with MPI 
communications.

Of concern for a GPU device implementation is the portability of a future 
IFS model where OpenACC parallelisation directives are applied to every 
DO loop nest, or handled by a DSL approach if that is pursued.

%% file: dwarf2.tex
\subsubsection{Scope}
\label{doc_dwarf2:sec:scope}
\textit{Bespoke scalable, preconditioned non-symmetric solvers} 
are one of the key areas of investigation in terms of Weather 
\& Climate dwarfs.
These arise in large time-step semi-implicit problems within 
compressible or sound-proof atmospheric models.
These solvers can be self-adaptive and may provide robust 
solution procedures for NWP applications. However, their 
scalability properties and effective preconditioning strategies 
must be explored.

Specifically, such solvers are suitable to be adopted in 
conjunction with compact-stencil schemes, such as the finite 
volume method, that are an attractive option for next generation 
HPC facilities given their reduced communication costs (due to 
their intrinsic data locality).

Dwarf-D-ellipticSolver-GCR implements a promising candidate, 
namely the Generalised Conjugate Residual (GCR) method. 
This approach aims to provide a robust solution procedure 
for NWP applications required by this key investigation 
area. In particular, the three-dimensional potential-flow 
problem constitutes an excellent benchmark to experiment 
with different preconditioners to evaluate its competitiveness, 
especially in terms of number of iterations required to 
reach convergence and scalability.

\subsubsection{Objectives}
\label{doc_dwarf2:sec:objectives}
The main objectives of this dwarf are to improve the computational 
time per iteration of the GCR algorithm and to reduce the number
of iterations to reach convergence. These two aspects are crucial,
since a consistent reduction of the computational cost might allow 
these methods, in conjunction with finite-volume discretizations,
to become very competitive for NWP applications, given their compact 
nature. This is particularly attractive for next-generation HPC 
infrastructures, while retaining a semi-implicit time discretization 
that allows large time-steps.

The evaluation of this dwarf will iclude:
\begin{enumerate}
\item Test the GCR approach on different hardware - especially 
GPUs and Accelerators - in order to identify the best solutions 
in terms of computational time per fixed number of iterations. 
Among the best solutions, it will also be necessary to identify 
the best compromise in terms of energy cost.
\item Test the GCR approach applied to the three-dimensional potential 
flow problem, i.e. \inlsh{fvm3d-potential} with different preconditioners 
in order to reduce the number of iterations to reach convergence.
\end{enumerate}

\subsubsection{Definition}
\label{doc_dwarf2:sec:definition}
Dwarf-D-ellipticSolver-GCR implements the solution of linear elliptic 
equations through the GCR method \cite{Eisenstat1983,Smolarkiewicz1994,
Smolarkiewicz2000}. 	
Specifically, we solve the potential flow equations that arise when 
discretizing atmospheric flows using a Finite Volume (FV) approach.

In particular, this dwarf solves a three-dimensional elliptic problem, 
namely a potential flow over a Gaussian-shaped hill on the sphere. 
In the following, we first describe the GCR method for a general 
linear elliptic problem and then we briefly outline the three-dimensional 
problem are available.

\paragraph{Generalized Conjugate-Residual approach}
In this section we follow closely appendix A of \cite{Thomas2003Spectral}.
In particular, we can formulate the GCR approach starting from the following 
linear elliptic problem:
\begin{equation}
\mathcal{L}(\psi) = \sum_{I=1}^{3}\frac{\partial}{\partial x^{I}}
\bigg(\sum_{J=1}^{3}\text{C}^{IJ}\frac{\partial \psi}{\partial x^{J}} 
+ \text{D}^{I}\bigg) - \text{A}\psi = \text{Q},
\label{doc_dwarf2:eq:linear-elliptic}
\end{equation}
where A, C$^{IJ}$, D$^{I}$, Q are variable coefficients and for which either 
periodic, Dirichlet or Neumann boundary conditions can be applied.
The discrete representation of a field is denoted by the subscript 
$i$ such that the discrete linear operator is represented by $\mathcal{L}_{i}$
while the inner product by $\langle\xi\,\zeta\rangle = \sum_{i}\xi_{i}\zeta_{i}$.

The GCR approach is formulated in terms of a preconditioner, denoted 
by $\mathcal{P}$, which is a linear operator that approximates $\mathcal{L}$ 
but which is easier to invert. This dwarf is intended to explore just 
left preconditioning strategies. These lead Eq.~\ref{doc_dwarf2:eq:linear-elliptic} 
to be substituted by the auxiliary problem $\mathcal{P}^{-1}[\mathcal{L}(\psi) 
- \text{Q}] = 0$. 

The GCR method presented in \cite{Eisenstat1983}, can then be obtained
through variational arguments (see for instance \cite{Smolarkiewicz1994,
Smolarkiewicz2000}). More specifically, the original problem, 
Eq.~\ref{doc_dwarf2:eq:linear-elliptic}, is augmented by the following $k$th-order 
damped oscillation equation:
\begin{equation}
\frac{\partial^{k}\mathcal{P}(\psi)}{\partial\tau^{k}} 
+ \frac{1}{\text{T}_{k-1}(\tau)}\frac{\partial^{k-1}\mathcal{P}(\psi)}{\partial\tau^{k-1}} 
+ \dots + \frac{1}{\text{T}_{1}(\tau)}\frac{\partial\mathcal{P}(\psi)}{\partial\tau} 
= \mathcal{L}(\psi) - \text{Q},
\label{doc_dwarf2:eq:damped}
\end{equation}
that is discretized in pseudo-time $\tau$.
This forms the affine discrete equation for the progression of the 
residual errors $r$. The final step is to identify the optimal parameters 
T$_{1}$, ..., T$_{k-1}$ and integration increment $\delta\tau$ that makes 
sure that the minimization of the residual errors in the norm defined 
by the inner product $\langle r\,r \rangle$ is successful.

\paragraph{Test case}
The dwarf solves a potential flow over a Gaussian-shaped hill. 
The governing equations of this problem may be written as follows:
\begin{equation}
\begin{array}{l}
\boldsymbol{v} = \boldsymbol{v}_{e} - \nabla\phi\\[0.5em]
\nabla\cdot(\rho\boldsymbol{v}) = 0
\end{array}
\label{doc_dwarf2:eq:potential}
\end{equation}  
where $\boldsymbol{v}_{e}$ is the ambient velocity which can be 
obtained from measurements or prescribed analytically and $\overline{\rho} = 
\overline{\rho}(z)$ is the reference density. In the above problem, 
we seek $\phi$ such that it satisfies the set of equations we aim 
to solve, thus providing a better initial guess for the given problem. 
Equation~\ref{doc_dwarf2:eq:potential} 
is cast in a non-orthogonal terrain-following systems of coordinates, 
$[x,y,z] = [x_{\text{cart}},y_{\text{cart}},H(z_{\text{cart}} - h)/(H - h]$,
with the subscript 'cart' referring to Cartesian coordinates, $H$
to the model depth and $h = h(x_{\text{cart}},y_{\text{cart}})$ 
to the mountain profile. 
The transformation of Eq.~\ref{doc_dwarf2:eq:potential} from the Cartesian 
frame of reference to the terrain-following frame of reference 
involves some metric terms that are fairly standard in atmospheric 
applications.
In particular, we can introduce the following transformation matrix
and Jacobian:
\begin{equation}
\begin{array}{l}
\displaystyle \text{G}^{IJ} = \sum_{K=1}^{2}\frac{\partial x^{I}}{\partial x_{\text{cart}}^{K}}
\frac{\partial x^{J}}{\partial x_{\text{cart}}^{K}}\\[0.5em]
\displaystyle \text{J} = \big[det(\text{G}^{IJ})\big]^{1/2},
\end{array}
\label{doc_dwarf2:eq:2d-transformations} 
\end{equation}
where the subscript `cart' denotes Cartesian coordinates.
For additional information the interested reader can refer 
to \cite{Smolarkiewicz1994}. 
Using these quantities, the original problem in Eq.~\ref{doc_dwarf2:eq:potential}
becomes:
\begin{equation}
\begin{array}{l}
\displaystyle u = u_{e} - \frac{\partial\phi}{\partial x} 
- \text{G}^{13}\frac{\partial\phi}{\partial z}\\[0.75em]
\displaystyle v = v_{e} - \frac{\partial\phi}{\partial y} 
- \text{G}^{23}\frac{\partial\phi}{\partial z}\\[0.75em]
\displaystyle w = w_{e} - \text{J}^{-1}\frac{\partial\phi}{\partial z}\\[0.75em]
\displaystyle \frac{\partial \rho^{*}u}{\partial x} + 
\frac{\partial \rho^{*}v}{\partial y} + 
\frac{\partial\rho^{*}\omega}{\partial z} = 0,
\end{array}
\label{doc_dwarf2:eq:transformed}
\end{equation}
where $\rho^{*} = \text{J}\overline{\rho}$ and the contravariant 
velocity $\omega$ is defined as $\omega = \text{J}^{-1}w + \text{G}^{13}u 
+ \text{G}^{23}v$. 

To obtain the associated boundary-value problem, we perform three steps.
We first substitute the $u$, $v$ and $w$ into the contravariant velocity 
$\omega$, we then substitute $u$, $v$ and the just obtained contravariant 
velocity $\omega$ into the last equation in Eq.~\ref{doc_dwarf2:eq:transformed}.
These first two steps lead to the following problem:
\begin{equation}
\nabla\cdot\Big(\rho^{*}\mathbf{P}_{f}(\phi,\nabla\phi)\Big),
\label{doc_dwarf2:eq:manipulated}
\end{equation}
where the components of the pressure force vector $\mathbf{P}_{f}$
are defined as follows:
\begin{equation}
\begin{array}{l}
\displaystyle \mathbf{p}_{f,x} = u\\[0.5em]
\displaystyle \mathbf{p}_{f,y} = v\\[0.5em]
\displaystyle \mathbf{p}_{f,z} = \omega_{e} - 
\text{G}^{13}\frac{\partial\phi}{\partial x} - 
\text{G}^{23}\frac{\partial\phi}{\partial y} - 
\text{G}^{33}\frac{\partial\phi}{\partial z}
\end{array}
\label{doc_dwarf2:eq:transformed}
\end{equation}
with $\omega_{e} = \text{J}^{-1}w_{e} + \text{G}^{13}u_{e} + \text{G}^{23}v_{e}$
and $\text{G}^{33} = (\text{G}^{13})^{2} + (\text{G}^{23})^{2} + 1/\text{J}^{2}$.
The third step involves multiplying Eq.~\ref{doc_dwarf2:eq:manipulated} 
by $-1/\rho^{*}$. This finally leads to the following equation:  
\begin{equation}
-\frac{1}{\rho^{*}}\nabla\cdot\bigg(\rho^{*}\boldsymbol{P}_{f}
(\phi,\nabla\phi)\bigg) = \frac{1}{\rho^{*}}\nabla\cdot\bigg(
\rho^{*}\big(\boldsymbol{v}^{*} - \boldsymbol{P}_{f}\big)\bigg) 
= 0,
\label{doc_dwarf2:eq:potential}
\end{equation}  
with the contravariant velocity vector $\boldsymbol{v}^{*}=\left(u,v,\omega\right)$. This equation (\ref{doc_dwarf2:eq:potential}) is solved using the GCR algorithm introduced above.
For additional details on this model problem refer to 
\cite{Smolarkiewicz1994}. Note that this is the target 
problem for this dwarf.

\newpage
\paragraph{Pseudo-algorithm}
The steps described in the previous subsections leads to the GCR 
pseudo-algorithm reported in algorithm~\ref{doc_dwarf2:alg:GCR}.
\begin{algorithm}[H]
\caption{Generalized Conjugate Residual (GCR) method}
\label{doc_dwarf2:alg:GCR}
\begin{algorithmic}
\vspace{.1cm}
\State For any initial guess $\psi_{i}^{0}$, \textbf{set}
\State $r_{i}^{0} = \mathcal{L}(\psi^{0}) - \text{Q}_{i}$
\State $p_{i}^{0} = \mathcal{P}^{-1}(r^{0})$, \textbf{then}:\\

\For{n = 1,2,...,until convergence}
 \For{$\nu$ = 0,1,...,k-1}
   \vspace{.4cm}
   \State $\displaystyle 
   \beta = -\frac{\langle r^{\nu}\,\mathcal{L}(p^{\nu})\rangle}{
   \mathcal{L}(p^{\nu})\,\mathcal{L}(p^{\nu})}$\\
   \State $\displaystyle 
   \psi^{\nu+1}_{i} = \psi^{\nu}_{i} + \beta\,p^{\nu}_{i}$\\
   \State $\displaystyle
   r^{\nu+1}_{i} = r^{\nu}_{i} + \beta\mathcal{L}_{i}(p^{\nu})$\\
   \State \textbf{if} $\Vert r^{\nu+1}\Vert \leq \epsilon$ \textbf{then} exit\\
   \State $\displaystyle  
   e_{i} = \mathcal{P}^{-1}_{i}(r^{\nu+1})$\\
   \State $\displaystyle
   \mathcal{L}_{i}(e) = \bigg[\sum_{I=1}^{3}\frac{\partial}{\partial x^{I}}
   \bigg(\sum_{J=1}^{3}\text{C}^{IJ}\frac{\partial e}{\partial x^{J}} + 
   \text{D}^{I}\,e\bigg) - A\,e\bigg]_{i}$\\
   \State $\displaystyle
   \textbf{for}\; \ell=0,...,\nu\;\;\;\alpha_{\ell} = 
   -\frac{\langle \mathcal{L}(e)\,\mathcal{L}(p^{\ell})\rangle}{\langle
   \mathcal{L}(p^{\ell})\,\mathcal{L}(p^{\ell})\rangle}$\\
   \State $\displaystyle
   p^{\nu+1}_{i} = e_{i} +     
   \sum_{\ell=0}^{\nu}\alpha_{\ell}\,p^{\ell}_{i}$\\
   \State $\displaystyle
   \mathcal{L}_{i}(p^{\nu+1}) = \mathcal{L}_{i}(e) + 
   \sum_{\ell=0}^{\nu}\alpha_{\ell}\,\mathcal{L}^{\ell}_{i}(p^{\ell})$
 \EndFor
 \State reset [$\psi$, $r$, $p$, $\mathcal{L}$]$_{i}^{k}$ 
 to [$\psi$, $r$, $p$, $\mathcal{L}$]$_{i}^{0}$
\EndFor
\end{algorithmic}
\end{algorithm}

\paragraph{I/O interfaces }
This dwarf has a simple input/output (I/O) layout. 
Specifically, the interfaces in terms of I/O data are as follows:
\begin{itemize}
\item \textbf{Input}: an Atlas-type field in grid-point space, defined 
on a Finite-Volume-type mesh. This represents the initial guess 
of the scalar function that is necessary to compute;
\item \textbf{Output}: an Atlas-type field in grid-point space, defined 
on a Finite-Volume-type mesh; This represents the converged 
value of the scalar function we needed to compute.
\end{itemize}
The dimensions of these Atlas-type fields  are determined 
by the grid employed that can be specified in the input files 
(.json format). Grids that are reasonable for current operations 
and for next generation global NWP are \textbf{TCo1279} 
(9 km global resolution -- currently operational), \textbf{TCo1999} 
(5 km global resolution), \textbf{TCo3999} (2.5 km global resolution), 
\textbf{TCo7999} (1.3 km global resolution). The acronym \textbf{TCo}
refers to the Cubic Octahedral grid.

To use these grids you need to change the \inlsh{grid} argument 
in the \inlsh{.json} files. For more details on how to run the dwarf, 
refer to section~\ref{doc_dwarf2:sec:run}.

\subsubsection{Prototypes}
\label{doc_dwarf2:sec:prototypes}
In this section we provide a brief description of each prototype 
implemented.

\paragraph{Prototype 1}
The first prototype, \inlsh{prototype1}, implements the elliptic solver core routines relying on the Atlas data-structure and 
based on the Fortran90 programming language.
The dwarf is divided into a main file and various module files:
\begin{itemize}
\item \inlsh{dwarf-D-ellipticSolver-GCR-prototype1.F90}, 
which implements the main program;
\item \inlsh{dwarf\_D\_ellipticSolver\_GCR\_coreLoop.F90},
which contains the core loops of the GCR algorithm as reported 
in algorithm~\ref{doc_dwarf2:alg:GCR};
\item \inlsh{dwarf\_D\_ellipticSolver\_GCR\_linear\_operators\_module.F90},
which contains the implementation of the linear operator $\mathcal{L}$;
\item \inlsh{dwarf\_D\_ellipticSolver\_GCR\_nabla\_operators\_module.F90},
which contains the implementation of the various `nabla' operators, 
such as Gradient, Divergence and Laplacian;
\item \inlsh{dwarf\_D\_ellipticSolver\_GCR\_preconditioner\_module.F90},
which contains a trivial preconditioner (this can be modified in order 
to build an efficient preconditioner);
\item \inlsh{dwarf\_D\_ellipticSolver\_GCR\_mappings\_module.F90},
which contains the necessary mappings for using spherical coordinates;
\item \inlsh{dwarf\_D\_ellipticSolver\_GCR\_topology\_module.F90},
which defines the surface topography of the problem;
\item \inlsh{dwarf\_D\_ellipticSolver\_GCR\_auxiliary\_module.F90},
which contains some support functionalities.
\end{itemize}
Note that the key part of the code defined in algorithm~\ref{doc_dwarf2:alg:GCR} 
is contained within the file \inlsh{dwarf\_D\_ellipticSolver\_GCR\_coreLoop.F90}, 
inside the subroutine \inlsh{GCR\_k}.

\subsubsection{Dwarf installation and testing}
\label{doc_dwarf2:sec:installation}
In this section we describe how to download and install 
the dwarf along with all its dependencies, and we show 
how to run it for a simple test case.

Note that dwarf-D-ellipticSolver-GCR is implemented using 
\textit{Atlas}, the ECMWF software framework that supports 
flexible data-structures for NWP. Currently, the dwarf 
is written in Fortran 2003. Extension to C++ can 
be envisioned if necessary and it can be implemented 
using \textit{Atlas}.

\paragraph{Download and installation}
The first step is to download and install the dwarf along 
with all its dependencies. With this purpose, it is possible 
to use the script provided under the ESCAPE software collaboration 
platform:\\
\url{https://git.ecmwf.int/projects/ESCAPE}.

Here you can find a repository called \inlsh{escape}.
You need to download it. There are two options to do this. One option is to use ssh. For this option you need to add an ssh key to your bitbucket account at \url{https://git.ecmwf.int/plugins/servlet/ssh/account/keys}. The link "SSH keys" on this website gives you instructions on how to generate the ssh key and add them to your account. Once this is done you should first create a 
folder named, for instance, ESCAPE, enter into it 
and subsequently download the repository by using the following the steps below:
\begin{lstlisting}[style=BashStyle]
mkdir ESCAPE
cd ESCAPE/
git clone ssh://git@git.ecmwf.int/escape/escape.git
\end{lstlisting}
The other option to download the repo is by using https instead of ssh. Instead of the git command above you then need to use 
\begin{lstlisting}[style=BashStyle]
git clone https://<username>@git.ecmwf.int/scm/escape/escape.git
\end{lstlisting}
where <username> needs to be replace by your bitbucket username.

Once the repository is downloaded into the \inlsh{ESCAPE} folder 
just created, you should find a new folder called \inlsh{escape}. 
The folder contains a sub-folder called \inlsh{bin} that has the 
python/bash script (called \inlsh{escape}) that needs to be 
run for downloading and installing the dwarf and its dependencies. 
To see the various options provided by the script you can type:
\begin{lstlisting}[style=BashStyle]
./escape/bin/escape -h
\end{lstlisting}
To download the dwarf you need to run 
the following command:
\begin{lstlisting}[style=BashStyle]
./escape/bin/escape checkout dwarf-D-ellipticSolver-GCR \ 
--ssh
\end{lstlisting}
To use https you need to replace --ssh with --user <username>. The commands above automatically check out the \inlsh{develop}
version of the dwarf. If you want to download a specific branch 
of this dwarf, you can do so by typing:
\begin{lstlisting}[style=BashStyle]
./escape/bin/escape checkout dwarf-D-ellipticSolver-GCR --ssh \
--version <branch-name>
\end{lstlisting}
An analogous approach can be used for the \inlsh{-\,-user} 
version of the command. You should now have a folder called 
\inlsh{dwarf-D-ellipticSolver-GCR}.

In the above command, you can specify several other optional 
parameters. To see all these options and how to use them you 
can type the following command:
\begin{lstlisting}[style=BashStyle]
./escape checkout -h
\end{lstlisting}

At this stage it is possible to install the dwarf 
and all its dependencies. This can be done in two 
different ways. The first way is to compile and 
install each dependency and the dwarf separately:
\begin{lstlisting}[style=BashStyle]
./escape/bin/escape generate_install dwarf-D-ellipticSolver-GCR
\end{lstlisting}
The command above will generate a script 
called \inlsh{install-dwarf-D-ellipticSolver-GCR} 
that can be run by typing:
\begin{lstlisting}[style=BashStyle]
./install-dwarf-D-ellipticSolver-GCR
\end{lstlisting}
This last step will build and install the dwarf 
along with all its dependencies in the following 
paths:
\begin{lstlisting}[style=BashStyle]
dwarf-D-ellipticSolver-GCR/builds/
dwarf-D-ellipticSolver-GCR/install/
\end{lstlisting}

The second way is to create a bundle that compiles 
and installs all the dependencies together:
\begin{lstlisting}[style=BashStyle]
./escape/bin/escape generate_bundle dwarf-D-ellipticSolver-GCR
\end{lstlisting}
This command will create an infrastructure to avoid
compiling the single third-party libraries individually
when some modifications are applied locally to one of 
them. To complete the compilation and installation process, 
after having run the above command for the bundle, simply 
follow the instructions on the terminal.

In the commands above that generate the installation 
file, you can specify several other optional parameters. 
To see all these options and how to use them you 
can type the following command:
\begin{lstlisting}[style=BashStyle]
./escape generate-install -h
./escape generate-bundle -h
\end{lstlisting}

\paragraph{Testing}
You should now verify that the dwarf works as expected.
For this purpose, we created a testing framework that
allows us to verify that the main features of the dwarf 
are working correctly.

In particular, for each sub-dwarf we provide various 
regression tests in order to allow the results to be 
consistent when the underlying algorithms are modified, 
and to test additional features or different hardware. 
The regression tests can be found in the folder \inlsh{test} 
that is located in each sub-dwarf folder. 
For each sub-dwarf we also provide scripts running the 
code on different architectures, e.g. the Cray HPC at 
ECMWF, that can be found in the folder \inlsh{run-scripts} 
located in each sub-dwarf folder. 
\begin{tipbox}
We encourage partners who are testing different architectures 
to add and/or modify the scripts!
\end{tipbox}

To run this verification, you should run the following 
command:
\begin{lstlisting}[style=BashStyle]
ctest -j<number-of-tasks>
\end{lstlisting}
from inside the \inlsh{builds/dwarf-D-ellipticSolver-GCR}
folder.
\begin{warningbox}
We strongly advise you to verify via ctest that 
the main functionalities of the dwarf are working 
properly any time you apply modifications to the 
code. Updates that do not pass the tests cannot 
be merged. 
In addition, if you add a new feature to the dwarf,
this should be supported by a test if the existing
testing framework is not already able to verify its
functionality.
\end{warningbox}
For instructions on how to run the executables 
see the next section.

\subsubsection{Run the Dwarf}
\label{doc_dwarf2:sec:run}
We first rename the following two paths for the sake 
of compactness:
\begin{lstlisting}[style=BashStyle]
srcs=dwarf-D-ellipticSolver-GCR/sources/ 
inst=dwarf-D-ellipticSolver-GCR/install/
\end{lstlisting}

To run the dwarf in your local machine, 
you could do so by using the executable files inside 
\begin{lstlisting}[style=BashStyle] 
inst/dwarf-D-ellipticSolver-GCR/bin/
\end{lstlisting}
In particular, the executables need the specification 
of a configuration file that can be found at
\begin{lstlisting}[style=BashStyle] 
sources/dwarf-D-ellipticSolver-GCR/config-files/
\end{lstlisting}
These is \inlsh{dwarf-D-ellipticSolver-GCR-O32.json}. 
and it specifies some parameters needed by the dwarf to run
properly.
 
In particular, this dwarf implements a potential flow 
problem over an idealised hill. To run it you should 
copy the \inlsh{dwarf-D-ellipticSolver-GCR-O32.json} 
file in the folder where you want to run the simulation 
(or alternatively specify its path on the command line). 
The executable can then be run as follows:
\begin{lstlisting}[style=BashStyle] 
inst/dwarf-D-ellipticSolver-GCR/bin/dwarf-D-ellipticSolver-GCR \
--config dwarf-D-ellipticSolver-GCR-O32.json
\end{lstlisting}
where, if the \inlsh{.json} file is not in the current 
directory, you can specify its path after \inlsh{-\,-config}.

If you instead want to run the dwarf on an HPC machine 
available to the ESCAPE partners, you can automatically 
generate the job submission script with the \inlsh{escape} 
file. More specifically, if you run the following command:
\begin{lstlisting}[style=BashStyle]
./escape generate-run -c \
inst/bin/dwarf-D-ellipticSolver-GCR \
--config dwarf-D-ellipticSolver-GCR-O32.json
\end{lstlisting}
This allows the code to generate the submission script 
for the given HPC machine you are targeting without submitting 
the actual job. The command above will in fact simply generate 
an \inlsh{escape.job} file in the current folder. This can
successively be submitted via \inlsh{qsub} on the HPC machine 
you want to run the simulation on.

In the above command you can specify several other optional 
parameters, such as wall-time, number of tasks, number of 
threads, etc. To see all these options and how to set them 
up you can type the following command:
\begin{lstlisting}[style=BashStyle]
./escape generate-run -h
\end{lstlisting}

\subsubsection{Integration}
This dwarf explores the solution of a linear elliptic operator
in spherical coordinates and that arises in the context of 
mesh-based discretizations (such as the finite volume method). 
In particular, this dwarf can be integrated in a global Weather 
\& Climate model involving a semi-implicit time-stepping scheme 
and a mesh-based spatial discretization. This, for instance, 
arises in the solution of the horizontal part of an NWP model.

So, the key aspects for this dwarf to be integrated within 
a Weather \& Climate model are:
\begin{itemize}
\item Mesh-based discretization (e.g. finite-volume or finite-element 
methods);
\item Solution of the 3D elliptic problem arising from the semi-implicit discretization in time where separability of horizontal and vertical coordinates is no longer satisfied (different from the semi-implicit discretization in existing spectral models).
\end{itemize}
If these three aspects are present in the NWP model, this dwarf 
can be integrated to solve the horizontal Laplacian of the model 
in a semi-implicit and mesh-based manner.

%% file: dwarf3.tex
\subsubsection{Scope}
The cloud and precipitation microphysics is an essential building block 
of any weather and climate prediction model as it is necessary to represent 
the effects of small-scale sub-grid physical processes, such as cloud 
and precipitation microphysics, through a parameterization of the grid-scale 
prognostic variables in the model.
The cloud microphysics scheme is a computationally expensive routine 
and alternative ways that can accelerate its computation will be extremely 
beneficial for the computational performance of weather and climate prediction 
models.

\subsubsection{Objectives}
The main objective of this dwarf is to assess the scalability limits 
of cloud microphysics, mainly based on the IFS scheme, on different 
hardware such as hosts (e.g. CPUs), devices (e.g. GPUs) and a hybrid 
combination of them. 
Some new implementations of the IFS-based cloud scheme are also envisioned.
These are believed to perform better on device-type hardware, thereby 
providing a more energy-efficient and computationally faster solution 
for the cloud scheme.

More specifically, we aim to investigate the following points:
\begin{itemize}
\item multi-core (e.g. Broadwell) and many-core hosts (e.g. Intel Knights Landing (KNL)) 
using multi-threading via OpenMP;
\item multi-device (e.g. multiple GPU nodes) implementing multi-threading 
using OpenACC, CUDA, etc;
\item explore different data-alignment / data-structure strategies 
to enhance compiler performance;
\item reduce dependency on other vertical levels by using data from previous time step 
that might provide finer grain parallelism (that would be beneficial 
for GPUs and KNL with OpenMP 4.5);
\end{itemize}
Each of the points outlined above will have a specific prototype 
implementation to permit a better understanding of the results, 
thus ultimately allowing the identification of the solution that 
guarantees the best compromise in terms of energy requirements 
/ time-to-solution.

\subsubsection{Definition of the Dwarf}
Weather and climate prediction models need to represent the 
effects of sub-grid scale physical processes. Radiation, turbulent 
mixing, convection, cloud and precipitation microphysics are 
examples of physical processes that are parametrized as a function 
of the grid-scale prognostic variables in models. This dwarf 
is the parametrizaton scheme for cloud and precipitation processes 
in the IFS, described by prognostic equations for cloud liquid water, 
cloud ice, rain, snow and a grid-box fractional cloud cover. The cloud 
scheme represents the sources and sinks of cloud and precipitation due 
to the major generation and destruction processes, including cloud formation 
by detrainment from cumulus convection, condensation, ice deposition, evaporation, 
hydrometeor collection, melting and freezing. The scheme is based on \cite{Tiedtke1993} 
but with an enhanced representation of the ice-phase in clouds and precipitation. 
A multi-dimensional implicit solver is used for the numerical solution of 
the cloud and precipitation prognostic equations.  A more detailed description 
of the formulation of the parametrization can be found in \cite{IFSdoc} with 
further discussion in \cite{ForbesandTompkins2011} and \cite{Forbesetal2011}.

\paragraph{Governing equations}
The equations for the tendency of the grid-box averaged cloud
liquid, cloud ice, rain and snow water contents are

\begin{equation}
\frac{\partial q_{\text{l}}}{\partial t} =Q(q_{\text{l}}) +S_{\text{conv}} 
+S_{\text{strat}}+S_{\text{melt}}^{\text{ice}}-S_{\text{dep}}^{\text{ice}}
-S_{\text{evap}}^{\text{liq}}-S_{\text{auto}}^{\text{rain}}-S_{\text {rime}}^{\text{snow}}
\end{equation}
\begin{equation}
\frac{\partial q_{\text{i}}}{\partial t} =Q(q_{\text{i}}) +S_{\text{conv}} 
+S_{\text{strat}}+S_{\text{dep}}^{\text{ice}}-S_{\text{melt}}^{\text{ice}}
-S_{\text{evap}}^{\text{ice}}-S_{\text{auto}}^{\text{snow}}
\end{equation}
\begin{equation}
\frac{\partial q_{\text{r}}}{\partial t} =Q(q_{\text{r}}) 
-S_{\text{evap}}^{\text{rain}}+S_{\text{auto}}^{\text{rain}}+S_{\text{melt}^{\text snow}} 
-S_{\text{frz}}^{\text{rain}}  
\end{equation}
\begin{equation}
\frac{\partial q_{\text{s}}}{\partial t} =Q(q_{\text{s}}) 
-S_{\text{evap}}^{\text{snow}}+S_{\text{auto}}^{\text{snow}}-S_{\text{melt}}^{\text{snow}} 
+S_{\text{frz}}^{\text{rain}}+S_{\text{rime}}^{\text{snow}}
\end{equation}

and for the cloud fraction,

\begin{equation}
\frac{\partial a}{\partial t} =Q(a) +\delta a_{\text{conv}} 
+\delta a_{\text{strat}}-\delta a_{\text{evap}}
\end{equation}

The terms on the right-hand side represent the following processes:
\begin{itemize}
\item    $Q(q), Q(a)$  -- rate of change of water contents and cloud
area due to transport through the boundaries of the grid volume (advection, 
sedimentation).
\item    $S_{\text{conv}}, \delta a_{\text{conv}}$ -- rate of formation 
of cloud water/ice and cloud area by convective processes.
\item    $S_{\text{strat}}, \delta a_{\text{strat}}$ -- rate of formation 
of cloud water/ice and cloud area by stratiform condensation processes.
\item    $S_{\text{evap}}$ -- rate of evaporation of cloud water/ice, 
rain/snow. 
\item    $S_{\text{auto}}$ -- rate of generation of precipitation from
cloud water/ice (autoconversion).
\item    $S_{\text{melt}}$ -- rate of melting ice/snow.
\item    $S_{\text{rime}}$ -- rate of riming (collection of cloud liquid 
drops).
\item    $S_{\text{frz}} $ -- rate of freezing of rain.
\item    $\delta a_{\text{evap}}$  -- rate of decrease of cloud area due 
to evaporation.
\end{itemize}

The large-scale budget equations for specific humidity $q_v$, 
and dry static energy $s =c_pT +gz$ in the cloud scheme are
\begin{equation}
\frac{\partial q_{\text{v}}}{\partial t} =Q(q_{\text{v}}) -S_{\text{strat}} +
S_{\text{evap}}
\end{equation}
and
\begin{equation}
\frac{\partial s}{\partial t} =Q(s)+L_{\text{vap}}(S_{\text{strat}} -
S_{\text{evap}})+L_{\text{fus}}(S_{\text{frz}} + S_{\text{rime}} - S_{\text{melt}})
\end{equation}
where $A(q_v)$ and $A(s)$ represent all processes except those
related to clouds, $L_{\text{vap}}$ is the latent heat
of condensation and $L_{\text{fus}}$ is the latent heat
of freezing.

Each of the microphysical source and sink terms is represented by an equation 
or set of equations that vary in complexity, from a simple linear form to more 
non-linear functions involving exponentials and power laws. Some terms are formulated
explicitly and others implicitly and they are combined in a multi-dimensional solver 
to produce the tendencies for the prognostic variables
(cloud liquid, cloud ice, rain, snow and humidity). Cloud fraction is treated
separately as this is a non-conservative variable. The temperature tendency due
to change in phase (vapour, liquid, ice) is calculated after the solver once the
final tendencies are known.

\paragraph{Integration of the equations}
  
The above equations governing the tendency for each prognostic cloud
variable within the cloud scheme can be written as:
\begin{equation}
\frac{\partial {q}_x}{\partial t} = A_x +
\frac{1}{\rho}\frac{\partial}{\partial z} \left( \rho V_x {q}_x \right)
\end{equation}
where $q_x$ is the specific water content for category $x$ (so $x=1$
represents cloud liquid, $x=2$ for rain, and so on), $A_x$ is the net source
or sink of $q_x$ through microphysical processes, and the last term
represents the sedimentation of $q_x$ with fall speed $V_x$.

The solution to this set of equations uses the upstream approach. Writing the
advection term in mass flux form and collecting all fast processes (relative 
to the model timestep) into an implicit term, gives:
\begin{equation}
\frac{q_x^{n+1}-q_x^{n}}{\Delta t} = A_x 
+\sum_{y=1}^{m}B_{xy}q_y^{n+1}
-\sum_{y=1}^{m}B_{yx}q_x^{n+1}
+\frac{ 
\rho_{k-1} V_{x} q_{x,k-1}^{n+1} - \rho V_{x} q_{x}^{n+1} }{ \rho \Delta
Z}
\label{doc_dwarf3:eqn-impl}
\end{equation}

for timestep $n$. The subscript "$k-1$" refers to a term calculated at the model
level above the present level $k$ for which all other terms are being
calculated. The matrix $\widetilde{B}$ (with terms $B_{xx}$, $B_{xy}$, $B_{yx}$)
represents all the implicit microphysical pathways such that $B_{xy}>0$
represents a sink of $q_y$ and a source of $q_x$. Matrix $\widetilde{B}$ is
positive-definite off the diagonal, with zero diagonal terms since $B_{xx}=0$ by
definition.  Some terms, such as the creation of cloud through condensation
resulting from adiabatic motion or diabatic heating, are more suitable for an
explicit framework, and are retained in the explicit term $A$.

For cloud fraction, there are no multi-dimensional dependencies, so the equation
simplifies to
\begin{equation}
\frac{a^{n+1}-a^{n}}{\Delta t} = A + B\,a^{n+1}
\end{equation}

However, for the cloud and precipitation variables,  a
matrix approach is required. Due to the cross-terms $q_y^{n+1}$, 
(\ref{doc_dwarf3:eqn-impl}) is rearranged to give a straight forward matrix equation which
can be solved with standard methods. 
The solution method is simplified by assuming the vertical advection terms due
to convective subsidence and sedimentation act only in the downward direction,
allowing the solution to be conducted level by level from the model top down.

The matrix on the left-hand side has the microphysical terms in isolation off the diagonal,
with the sedimentation term on the diagonal, thus the matrix equation for a 3-variable 
system is
{\footnotesize
\begin{eqnarray}
\left(
\begin{array}{ccc}
1+ \Delta t(\frac{V_1}{\Delta z} + B_{21} + B_{31}) & -\Delta t B_{12}
& -\Delta t B_{13} \\
-\Delta t B_{21} & 1+ \Delta t(\frac{V_2}{\Delta z} + B_{12} + B_{32}) &
-\Delta t B_{23}  \\
-\Delta t B_{31} & -\Delta t B_{32} & 1+ \Delta t(\frac{V_3}{\Delta z} 
+ B_{13} + B_{23}) \\
\end{array}
\right) \cdot
\left(
\begin{array}{c}
q_1^{n+1} \\
q_2^{n+1} \\
q_3^{n+1} \\
\end{array}
\right)
= \nonumber \\
\newline
\left[ 
q_1^{n} + \Delta t \left(A_1 +
\frac{ \rho_{k-1} V_{1} q_{1,k-1}^{n+1}}{\rho \Delta Z} \right) 
,q_2^{n} + \Delta t \left(A_2 +
\frac{ \rho_{k-1} V_{2} q_{2,k-1}^{n+1}}{\rho \Delta Z} \right) 
,q_3^{n} + \Delta t \left(A_3 +
\frac{ \rho_{k-1} V_{3} q_{3,k-1}^{n+1}}{\rho \Delta Z} \right) 
\right].
\end{eqnarray}
}

There are some aspects that require attention.  Firstly, although implicit terms are unable 
to reduce a cloud category to zero, the explicit can, and often will, achieve this.  Thus 
safety checks are required to ensure that all end-of-timestep variables remain positive 
definite, in addition to ensuring conservation.  Practically, to aid the conservation requirement, 
the explicit source and sink terms are thus also generalised from a vector $\vec A$ to an 
anti-symmetric matrix  $\widetilde{A}$,
\begin{eqnarray}
\widetilde{A}=\left(
\begin{array}{ccc}
A_{11} & A_{21} & A_{31} \\
-A_{12} & A_{22} & A_{32} \\
-A_{13} & -A_{23} & A_{33} \\
\end{array}
\right)
\end{eqnarray}
Thus $A_{xy}>0$ represents a source of $q_{x}$ and a sink of $q_{y}$, and the original vector 
for $A$ can be obtained by summing over the rows. Although this matrix approach involves 
a degree of redundancy, it is a simple method of ensuring conservation properties. The matrix 
diagonals $A_{xx}$ contain the 'external' sources of $q_{x}$ such as the cloud water detrainment 
terms from the convection scheme.

In order to simultaneously guarantee conservation and positive-definite properties, the sum of all 
sinks for a given variable are scaled to avoid negative values.

\newpage
\paragraph{Pseudo-algorithm}
The steps described in the previous subsections leads to the cloud microphysics pseudo-algorithm 
reported in the following blocks of code. Note that, to solve the cloud/precipitation/vapour calculations, 
(see also Algorithm~\ref{doc_dwarf3:alg:core-loop}), this dwarf uses an LU decomposition 
with non-pivoting recursive factorization followed by back substitution to give the cloud variables 
at the new timestep $q_{n+1}$.

\newpage
\begin{algorithm}[H]
\caption{Core cloud micro physics loop}\label{doc_dwarf3:alg:core-loop}
\begin{algorithmic}
\State \textbf{set} constants
\vspace{.1cm}
\State \textbf{for} jk = ncldtop,...,klev
\Indent
\LineComment{initialize variables}
\Indent
\State $A_{a} = 0$;\;\;\; $B_{a} = 0$; \;\;\; 
$A_{q_{x} q_{y}} = 0$; \;\;\; $B_{q_{x} q_{y}} = 0$
\EndIndent
\vspace{.2cm}
\State \textcolor{blue}{\textbf{compute} cloud processes} \label{doc_dwarf3:lis:cloud-proc}
\State \textcolor{blue}{\textbf{compute} precipitation processes} \label{doc_dwarf3:lis:precipitation-proc}
\vspace{.3cm}
\State \textbf{solve} for cloud cover (a) at $t^{n+1}$: $a^{n+1} = (a^{n} + A_{a})/(1 + B_{a})$
\vspace{.1cm}
\LineComment{cloud microphysics calculations}
\Indent
\vspace{-.2cm}
\State ---------------------------------------------------------------------------------------
\vspace{-.2cm}
\For{jl = 0,...,ngptot}
\For{jm = 0,...,mtot}
\For{jn = 0,...,ntot}
\State \textbf{add up} all sink terms and \textbf{calculate} overshoot
\State \textbf{if} overshoot == false \textbf{then} scaling factor = 1
\State \textbf{else if} overshoot == true \textbf{then} compute scaling factor 
\State \textbf{scale} sink terms in the correct order
\EndFor
\EndFor
\EndFor
\vspace{-.2cm}
\State ---------------------------------------------------------------------------------------
\vspace{-.2cm}
\EndIndent
\vspace{.5cm}
\LineComment{cloud/precipitation/vapour calculations}
\Indent
\vspace{-.2cm}
\State ---------------------------------------------------------------------------------------
\vspace{-.2cm}
\For{jl = 0,...,ngptot}
\For{jm = 0,...,mtot}
\For{jn = 0,...,ntot}
\If{jn == jm}
\State zqlhs(jl,jn,jm) = 1 + zfallsink(jl,jm)
\For{jo = 0,...,nclv}
\State $zqlhs(jl,jn,jm)=zqlhs(jl,jn,jm) + zsolqb(jl,jo,jn)$ 
\EndFor
\EndIf
\If{jn != jm}
\State $zqlhs(jl,jn,jm)= -zsolqb(jl,jn,jm)$ 
\EndIf
\EndFor
\EndFor
\EndFor
\State ....
\EndIndent
\EndIndent
\algstore{bkbreak1}
\end{algorithmic}
\end{algorithm}
\begin{algorithm}[H]
\begin{algorithmic}
\algrestore{bkbreak1}
\Indent
\Indent
\State ...
\LineComment{continuing cloud/precipitation/vapour calculations}
\LineComment{Set the right hand side of the equation (explicit terms)}
\State $q_{n+1} = q_{n} + A_{q_{x} q_{y}}$
\vspace{.1cm}
\State \textbf{solve} matrix equation (11) \label{doc_dwarf3:lst:equation11} 
\State \textbf{compute} precipitation fluxes (loops over jl and jm)
\vspace{.2cm}
\State \textbf{update} tendencies (jl,jm) - temperature and cloud condensate
\Indent
\State \textbf{compute} temperature changes from phase changes
\State \textbf{compute} cloud tendencies $q_{n+1} - q_{n}$
\EndIndent
\vspace{.1cm}
\State \textbf{update} tendencies (jl) - humidity and cloud cover
\vspace{-.2cm}
\EndIndent
\State ---------------------------------------------------------------------------------------
\EndIndent
\State \textbf{end for} (close main loop over levels jk)
\State \textbf{compute} flux changes for diagnostics (jl,jk-loops)
\end{algorithmic}
\end{algorithm}

\newpage
\begin{algorithm}[H]
\caption{Computation of cloud processes (see line \ref{doc_dwarf3:lis:cloud-proc} 
in Algorithm~\ref{doc_dwarf3:alg:core-loop})}
\label{doc_dwarf3:alg:cloud-proc}
\begin{algorithmic}
\vspace{.2cm}
\LineComment{supersaturation due to change in humidity}
\Indent
\vspace{-.2cm}
\State ---------------------------------------------------------------------------------------
\vspace{-.2cm}
\State - from this timestep (jl-loop)
\State $A_{q_li q_v} = A_{q_li q_v} + S_{q_{zsupsat}}$
\State $A_{q_v q_li} = A_{q_v q_li} - S_{q_{zsupsat}}$
\State $A_a = A_a + S_{a_zsupsat}$
\State - from previous timestep (jl-loop)
\State $A_{q_l q_l} = A_{q_l q_l} + S_{q_{psupsat}}$
\State $A_{q_i q_i} = A_{q_i q_i} + S_{q_{psupsat}}$
\State $A_a = A_a + S_{a_psupsat}$
\vspace{-.2cm}
\State ---------------------------------------------------------------------------------------
\vspace{.4cm}
\EndIndent

\LineComment{detrain cloud from convection (jl-loop)}
\Indent
\vspace{-.2cm}
\State ---------------------------------------------------------------------------------------
\vspace{-.2cm}
\State $A_{q_l q_l} = A_{q_l q_l} + S_{q_{zconvsrce}}$
\State $A_{q_i q_i} = A_{q_i q_i} + S_{q_{zconvsrce}}$
\vspace{-.2cm}
\State ---------------------------------------------------------------------------------------
\vspace{.4cm}
\EndIndent

\LineComment{environmental subsidence and evaporation (jl,jm-loops)}
\Indent
\vspace{-.2cm}
\State ---------------------------------------------------------------------------------------
\vspace{-.2cm}
\State - subsidence source (explicit - dependency on level above)
\State $A_{q_l q_l} = A_{q_l q_l} + S_{q_{zlcust}}$
\State $A_{q_i q_i} = A_{q_i q_i} + S_{q_{zlcust}}$
\State $A_a = A_a + S_{a_zacust}$
\State - evaporation (explicit)    
\State $A_{q_li q_v} = A_{q_li q_v} + S_{q_{zevap}}$
\State $A_{q_v q_li} = A_{q_v q_li} - S_{q_{zevap}}$
\State - Subsidence sink (implicit)
\State $B_{q_l q_l} = B_{q_l q_l} + S_{q_{zmfdn}}$
\State $B_{q_i q_i} = B_{q_i q_i} + S_{q_{zmfdn}}$
\State $B_a = B_a + S_{a_zmfdn}$
\vspace{-.2cm}
\State ---------------------------------------------------------------------------------------
\vspace{.4cm}
\EndIndent

\LineComment{erosion of clouds by turbulent mixing (jl-loop)}
\Indent
\vspace{-.2cm}
\State ---------------------------------------------------------------------------------------
\vspace{-.2cm}
\State $A_{q_li q_v} = A_{q_li q_v} + S_{q_{zleros}}$
\State $A_{q_v q_li} = A_{q_v q_li} - S_{q_{zleros}}$
\State $A_a = A_a + S_{a_zaeros}$
\vspace{-.2cm}
\State ---------------------------------------------------------------------------------------
\vspace{.4cm}
\EndIndent
\State ...
\algstore{bkbreak2}
\end{algorithmic}
\end{algorithm}
\begin{algorithm}[H]
\begin{algorithmic}
\algrestore{bkbreak2}
\vspace{.2cm}
\State ...
\vspace{.2cm}
\LineComment{condensation/evaporation (jl-loop)}
\Indent
\vspace{-.2cm}
\State ---------------------------------------------------------------------------------------
\vspace{-.2cm}
\State \textbf{call cuadjtq} to calculate saturation adjustment (jl,jk 2d arrays)
\State - evaporation of cloud
\State $A_{q_li q_v} = A_{q_li q_v} + S_{q_{zlevap}}$
\State $A_{q_v q_li} = A_{q_v q_li} - S_{q_{zlevap}}$
\State - condensation in existing cloud
\State $A_{q_li q_v} = A_{q_li q_v} + S_{q_{zlcond1}}$
\State $A_{q_v q_li} = A_{q_v q_li} - S_{q_{zlcond1}}$
\State - condensation of new cloud
\State $A_{q_li q_v} = A_{q_li q_v} + S_{q_{zlcond2}}$
\State $A_{q_v q_li} = A_{q_v q_li} - S_{q_{zlcond2}}$
\State $A_a = A_a + S_{a_zacond}$
\vspace{-.2cm}
\State ---------------------------------------------------------------------------------------
\vspace{.4cm}
\EndIndent
\LineComment{growth of ice by vapour deposition (jl-loop)}
\Indent
\vspace{-.2cm}
\State ---------------------------------------------------------------------------------------
\vspace{-.2cm}
\State $A_{q_i q_l} = A_{q_i q_l} + S_{q_{zdepos}}$
\State $A_{q_l q_i} = A_{q_l q_i} + S_{q_{zdepos}}$
\vspace{-.2cm}
\State ---------------------------------------------------------------------------------------
\EndIndent
\end{algorithmic}
\end{algorithm}

\newpage
\begin{algorithm}[H]
\caption{Computation of precipitation processes (see line \ref{doc_dwarf3:lis:precipitation-proc} 
in Algorithm~\ref{doc_dwarf3:alg:core-loop})}
\label{doc_dwarf3:alg:cloud-proc}
\begin{algorithmic}
\vspace{.2cm}
\LineComment{sedimentation of ice, rain, snow (jl,jm-loops, dependency on level above)}
\Indent
\vspace{-.2cm}
\State ---------------------------------------------------------------------------------------
\vspace{-.2cm}
\State $A_{q_i q_i} = A_{q_i q_i} + S_{q_{zfallsrce}}$
\State $A_{q_r q_r} = A_{q_r q_r} + S_{q_{zfallsrce}}$
\State $A_{q_s q_s} = A_{q_s q_s} + S_{q_{zfallsrce}}$
\State \textbf{update} precipitation cover zcovptot and related variables
\vspace{-.2cm}
\State ---------------------------------------------------------------------------------------
\vspace{.4cm}
\EndIndent

\LineComment{autoconversion of ice to snow (implicit) (jl-loop)}
\Indent
\vspace{-.2cm}
\State ---------------------------------------------------------------------------------------
\vspace{-.2cm}
\State $B_{q_s q_i} = B_{q_s q_i} + S_{q_{zsnowaut}}$
\vspace{-.2cm}
\State ---------------------------------------------------------------------------------------
\vspace{.4cm}
\EndIndent

\LineComment{autoconversion of liquid to rain (jl-loop)}
\Indent
\vspace{-.2cm}
\State ---------------------------------------------------------------------------------------
\vspace{-.2cm}
\State $A_{q_rs q_l} = A_{q_rs q_l} + S_{q_{zrainaut}}$
\State $A_{q_l q_rs} = A_{q_l q_rs} - S_{q_{zrainaut}}$
\vspace{-.2cm}
\State ---------------------------------------------------------------------------------------
\vspace{.4cm}
\EndIndent

\LineComment{riming - collection of cloud liquid by snow (implicit) (jl-loop,}
\State dependency on level above)
\Indent
\vspace{-.2cm}
\State ---------------------------------------------------------------------------------------
\vspace{-.2cm}
\State $B_{q_s q_l} = B_{q_s q_l} + S_{q_{zsnowrime}}$
\vspace{-.2cm}
\State ---------------------------------------------------------------------------------------
\vspace{.4cm}
\EndIndent
\LineComment{melting of snow and ice (jl,jm-loops, dependency on level above)}
\Indent
\vspace{-.2cm}
\State ---------------------------------------------------------------------------------------
\vspace{-.2cm}
\State $A_{q_r q_is} = A_{q_r q_is} + S_{q_{zmelt}}$
\State $A_{q_is q_r} = A_{q_is q_r} - S_{q_{zmelt}}$
\vspace{-.2cm}
\State ---------------------------------------------------------------------------------------
\vspace{.4cm}
\EndIndent
\LineComment{freezing of rain (jl-loop, dependency on level above)}
\Indent
\vspace{-.2cm}
\State ---------------------------------------------------------------------------------------
\vspace{-.2cm}
\State $A_{q_s q_r} = A_{q_s q_r} + S_{q_{zfrz}}$
\State $A_{q_r q_s} = A_{q_r q_s} - S_{q_{zfrz}}$
\vspace{-.2cm}
\State ---------------------------------------------------------------------------------------
\vspace{.4cm}
\EndIndent
\LineComment{freezing of cloud liquid (jl-loop)}
\Indent
\vspace{-.2cm}
\State ---------------------------------------------------------------------------------------
\vspace{-.2cm}
\State $A_{q_i q_l} = A_{q_i q_l} + S_{q_{zfrz}}$
\State $A_{q_l q_i} = A_{q_l q_i} - S_{q_{zfrz}}$
\vspace{-.2cm}
\State ---------------------------------------------------------------------------------------
\vspace{.4cm}
\EndIndent
\State ...
\algstore{bkbreak3}
\end{algorithmic}
\end{algorithm}
\begin{algorithm}[H]
\begin{algorithmic}
\algrestore{bkbreak3}
\vspace{.2cm}
\State ...
\vspace{.2cm}
\LineComment{rain evaporation (jl-loop, dependency on level above)}
\Indent
\vspace{-.2cm}
\State ---------------------------------------------------------------------------------------
\vspace{-.2cm}
\State $A_{q_v q_r} = A_{q_v q_r} + S_{q_{zevap}}$
\State $A_{q_r q_v} = A_{q_r q_v} - S_{q_{zevap}}$
\vspace{-.2cm}
\State ---------------------------------------------------------------------------------------
\vspace{.4cm}
\EndIndent
\LineComment{snow evaporation (jl-loop, dependency on level above)}
\Indent
\vspace{-.2cm}
\State ---------------------------------------------------------------------------------------
\vspace{-.2cm}
\State $A_{q_v q_s} = A_{q_v q_s} + S_{q_{zevap}}$
\State $A_{q_s q_v} = A_{q_s q_v} - S_{q_{zevap}}$
\vspace{-.2cm}
\State ---------------------------------------------------------------------------------------
\vspace{.4cm}
\EndIndent
\end{algorithmic}
\end{algorithm}

\newpage
\paragraph{I/O interfaces}
The primary I/O interface data is as follows:
\begin{itemize}
\item \textbf{Input}: an array of vertical profiles at the start of 
the timestep in grid point space of temperature, humidity, cloud liquid water, cloud ice, cloud fraction, rain 
and snow, as well as the accumulated tendencies for these variables from all the 
processes up to this point (dynamics and physics). Adding the accumulated tendencies 
to the start of timestep values gives the updated values at the current point in the 
timestep.
\item \textbf{Output}: an array of vertical profiles of the tendencies of temperature, 
humidity, cloud liquid water, cloud ice, cloud fraction, rain and snow calculated 
within this call of the cloud scheme.
\end{itemize}
The dimensions of the grid point fields are (KLON,KLEV) where KLON is the number 
of grid columns and KLEV is the number of levels. KLEV is 91 or 137 for current operational 
applications and could be around 200 for next generation systems. Grid columns are 
completely independent of each other for the call to the cloud scheme, so KLON can 
be any value depending on the domain decomposition. For the full globe for current 
operations and for next generation NWP, typical global number of grid columns are 
of order 6.E6 for TCo1279 (9 km global resolution ? currently operational), 2.E7 for 
TCo1999 (5 km global resolution), 8.E7 for TCo3999 (2.5 km global resolution), and 
3.E8 for TCo7999 (1.3 km global resolution). The acronym TCo refers to the Cubic 
Octahedral grid.

\subsubsection{Prototypes}
In this section we describe the prototypes available.

\paragraph{Prototype1}
The first prototype implements the cloud microphysics using 
the column-based scheme coming from the IFS. The implementation 
is optimised to run on a traditional host-based machine.

\subsubsection{Dwarf installation and testing}
In this section we describe how to download and install 
the dwarf along with all its dependencies and we show 
how to run it for a simple test case.

\paragraph{Download and installation}
The first step is to download and install the dwarf along 
with all its dependencies. With this purpose, it is possible 
to use the script provided under the ESCAPE software collaboration 
platform:\\
\url{https://git.ecmwf.int/projects/ESCAPE}.

Here you can find a repository called \inlsh{escape}.
You need to download it. There are two options to do this. One option is to use ssh. For this option you need to add an ssh key to your bitbucket account at \url{https://git.ecmwf.int/plugins/servlet/ssh/account/keys}. The link "SSH keys" on this website gives you instructions on how to generate the ssh key and add them to your account. Once this is done you should first create a 
folder named, for instance, ESCAPE, enter into it 
and subsequently download the repository by using the following the steps below:
\begin{lstlisting}[style=BashStyle]
mkdir ESCAPE
cd ESCAPE/
git clone ssh://git@git.ecmwf.int/escape/escape.git
\end{lstlisting}
The other option to download the repo is by using https instead of ssh. Instead of the git command above you then need to use 
\begin{lstlisting}[style=BashStyle]
git clone https://<username>@git.ecmwf.int/scm/escape/escape.git
\end{lstlisting}
where <username> needs to be replace by your bitbucket username.

Once the repository is downloaded into the \inlsh{ESCAPE} folder 
just created, you should find a new folder called \inlsh{escape}. 
The folder contains a sub-folder called \inlsh{bin} that has the 
python/bash script (called \inlsh{escape}) that needs to be 
run for downloading and installing the dwarf and its dependencies. 
To see the various options provided by the script you can type:
\begin{lstlisting}[style=BashStyle]
./escape/bin/escape -h
\end{lstlisting}
To download the dwarf you need to run 
the following command:
\begin{lstlisting}[style=BashStyle]
./escape/bin/escape checkout dwarf-P-cloudMicrophysics-IFSScheme \ 
--ssh
\end{lstlisting}
To use https you need to replace --ssh with --user <username>. The commands above automatically check out the \inlsh{develop}
version of the dwarf. If you want to download a specific branch 
of this dwarf, you can do so by typing:
\begin{lstlisting}[style=BashStyle]
./escape/bin/escape checkout dwarf-P-cloudMicrophysics-IFSScheme --ssh \
--version <branch-name>
\end{lstlisting}
An analogous approach can be used for the \inlsh{-\,-user} 
version of the command. You should now have a folder called 
\inlsh{dwarf-P-cloudMicrophysics-IFSScheme}.

In the above command, you can specify several other optional 
parameters. To see all these options and how to use them you 
can type the following command:
\begin{lstlisting}[style=BashStyle]
./escape checkout -h
\end{lstlisting}

At this stage it is possible to install the dwarf 
and all its dependencies. This can be done in two 
different ways. The first way is to compile and 
install each dependency and the dwarf separately:
\begin{lstlisting}[style=BashStyle]
./escape/bin/escape generate-install dwarf-P-cloudMicrophysics-IFSScheme
\end{lstlisting}
The command above will generate a script 
called \inlsh{install-dwarf-P-cloudMicrophysics-IFSScheme} 
that can be run by typing:
\begin{lstlisting}[style=BashStyle]
./install-dwarf-P-cloudMicrophysics-IFSScheme
\end{lstlisting}
This last step will build and install the dwarf 
along with all its dependencies in the following 
paths:
\begin{lstlisting}[style=BashStyle]
dwarf-P-cloudMicrophysics-IFSScheme/builds/
dwarf-P-cloudMicrophysics-IFSScheme/install/
\end{lstlisting}

The second way is to create a bundle that compile 
and install all the dependencies together:
\begin{lstlisting}[style=BashStyle]
./escape/bin/escape generate-bundle dwarf-P-cloudMicrophysics-IFSScheme
\end{lstlisting}
This command will create an infrastructure to avoid
compiling the single third-party libraries individually
when some modifications are applied locally to one of 
them. To complete the compilation and installation process, 
after having run the above command for the bundle, simply 
follow the instructions on the terminal.

In the commands above that generate the installation 
file, you can specify several other optional parameters. 
To see all these options and how to use them you 
can type the following command:
\begin{lstlisting}[style=BashStyle]
./escape generate-install -h
./escape generate-bundle -h
\end{lstlisting}

\paragraph{Testing}
Not supported yet.



\subsubsection{Run the dwarf}
For the sake of compactness, we rename the various main 
folders of the downloaded dwarf as follows 
\begin{lstlisting}[style=BashStyle]
srcs=dwarf-P-cloudMicrophysics-IFSScheme/sources/
inst=dwarf-P-cloudMicrophysics-IFSScheme/sources/
\end{lstlisting}
To run a simple test case you need to link the input file, 
called \inlsh{cloudsc.bin}, required by this dwarf to the 
directory from where you intend to run the excutable.
This is located in the \inlsh{config-files} subfolder inside 
the \inlsh{dwarf-P-cloudMicrophysics-IFSScheme} directory 
and the link can be created as follows
\begin{lstlisting}[style=BashStyle]
ln -s srcs/dwarf-P-cloudMicrophysics-IFSScheme/config-files/cloudsc.bin .
\end{lstlisting}
The input data \inlsh{cloudsc.bin} is a Fortran unformatted stream binary 
(no record delimiters). It contains data for just 100 grid point 
columns and will be inflated to full spectre of NGPTOT, where 
NGPTOT is the number of grid point columns. To run a simple 
test you can now type:
\begin{lstlisting}[style=BashStyle]
inst/bundle/bin/dwarf-P-cloudMicrophysics-IFSScheme-prototype1 \
OMP NGPTOT NPROMA-list
\end{lstlisting}
where OMP is the number of threads for OMP-parallel regions, 
NGPTOT is the already mentioned number of grid point columns, 
NPROMA-list is a list of NPROMAs to use. For a simple test-case 
you can try the following parameter-combination:
\begin{lstlisting}[style=BashStyle]
inst/bin/dwarf-P-cloudMicrophysics-IFSScheme-prototype1 4 160000 2
\end{lstlisting}
Note that if you installed the dwarf through the bundle option, 
the executable can be run with the following command:
\begin{lstlisting}[style=BashStyle]
inst/bundle/bin/dwarf-P-cloudMicrophysics-IFSScheme-prototype1 4 160000 2
\end{lstlisting}

\subsubsection{Integration}
The cloud scheme is representative of the physical processes required 
by any NWP model. Its integration is therefore naturally 
guaranteed independently from the dynamical core being 
used.

%% file: dwarf4.tex
\subsubsection{Scope}

\par
Spectral methods have obvious advantages in atmospheric models. First, they provide a very \emph{accurate} calculation of spatial derivatives. Second, solving elliptic partial differential equations is almost trivial in spectral space. Therefore, spectral methods allow for implicit timestepping methods, which in turn permits to take large timesteps. This means that fewer timesteps are needed to reach a predefined forecast range. In this sense, spectral methods also help in the \emph{efficiency} of an atmospheric model. For these reasons, spectral methods can be considered as a reference to which alternatives can be compared.
\par
At the heart of spectral methods lies the decomposition of a spatial field into harmonic functions, i.e. eigenfunctions of the Laplacian operator. For global models, the harmonic functions are spherical harmonics, which consist of a product of harmonic functions in the zonal direction and associated Legendre polynomials in the meridional direction. For limited area models (LAMs) with a rectangular domain, the harmonic functions are somewhat simpler, and consist of a product of harmonic functions in the zonal and meridional directions (hence \emph{bi}-Fourier).
\par
The spectral transform on a spherical domain is implemented in another dwarf
(dwarf-D-spectralTransform-sphericalHarmonics). The dwarf presented in this document (dwarf-D-spectralTransform-BiFourier)
provides the LAM equivalent of dwarf-D-spectralTransform-sphericalHarmonics. As such, dwarf-D-spectralTransform-BiFourier will supply an essential building block of spectral LAMs such as the ALADIN system \cite{Benard2010}. This dwarf will allow to test the porting, scalability and energy-efficiency of spectral methods in a LAM context on the next generation of HPC machines with heterogeneous hardware components. Especially the implementation of this dwarf on optical co-processors (as developed by \emph{Optalysys}, partner in the ESCAPE project) is promising.

\subsubsection{Objectives}
The main objective of this dwarf is to test the efficiency of LAM spectral 
transforms (i.e. Bi-Fourier) on emerging hardware. More specifically, 
the aim is to port the underlying code to an accelerator or many-core 
architecture (these will be also referred to as {\bfseries devices}) 
environment and try to match or exceed the performance that 
can today be achieved on conventional multi-core (also referred 
to as {\bfseries host}) systems such as the CRAY XC-30 at ECMWF. 
In particular, it is important to achieve a time-to-solution comparable 
or better than the current implementation on host systems while saving 
energy due to the use of accelerator devices.

The detailed goals are therefore:
\begin{itemize}
\item to measure the time-to-solution provided by implementations 
of the spectral transform on different hardware, 
\item to measure the energy-to-solution and 
\item to find the best compromise that minimises both the time-to-solution 
and the energy-to-solution.
\end{itemize}
To achieve these goals we currently provide one prototype that implements 
the dwarf without using the Atlas  data-structure. This prototype currently 
support only standard host architectures and multi-threading is achieved 
through the use of OpenMP. The development of at least two other prototypes 
is essential to test device-type architectures and the Optalysis optical processors.

\subsubsection{Definition of the Dwarf}
\paragraph{Bi-Fourier spectral transform}
\par
Dwarf-D-spectralTransform-BiFourier implements the spectral transform method on a rectangular domain. This comprises
transforms between gridpoint space, where a field is represented by values in the domain gridpoints, and spectral space,
where a field is represented by the amplitudes of the composing harmonic functions. These spectral transforms are
performed consecutively in the zonal and meridional directions. In each of these directions, the transforms are Fast
Fourier transforms (FFTs), equivalent to the zonal transforms of dwarf-D-spectralTransform-sphericalHarmonics.
\par
The Fourier transform used in this dwarf is defined as follows:
\begin{itemize}
	\item Direct transform (i.e. gridpoint space to spectral space):
		\begin{align}
			a_k	&=\frac{1}{n}\sum_{j=0}^{n-1}	\psi_j \cos\left(\frac{2jk\pi}{n}\right)	\\
			b_k	&=-\frac{1}{n}\sum_{j=0}^{n-1}	\psi_j \sin\left(\frac{2jk\pi}{n}\right)
		\end{align}
		where $a_k$ and $b_k$ are the amplitudes of the harmonic functions with wavenumber $k$, $\psi_j$ is the value of the field in gridpoint $j$, and $n$ is the number of gridpoints in the zonal or meridional direction.
	\item Inverse transform (i.e. spectral space to gridpoint space):
		\begin{equation}
			\psi_j=\sum_{k=0}^{n-1}c_k \exp\left(\frac{2ijk\pi}{n}\right)
		\end{equation}
		where $c_k=a_k+ib_k$ for $k< n/2$, and $c_k=a_{n-k}-ib_{n-k}$ for $k\geq n/2$.
\end{itemize}
The pseudo-algorithm underlying the definition just specified is reported in section~\ref{doc_dwarf4:sec:pseudo}.

\paragraph{Pseudo-algorithm}
\label{doc_dwarf4:sec:pseudo}
\par
The current implementation of dwarf-D-spectralTransform-BiFourier is based on the setup 
and FFT routines of IFS/ARPEGE/ALADIN cycle 43 and is encapsulated in \inlsh{prototype1}. 
This implementation is based on the recursive Cooley-Tukey algorithm, which has a computational 
complexity of $\mathcal{O}(N\log N)$. The implementation requires the number of gridpoints 
to be factorizable into powers of $2$, $3$ or $5$: $n=2^{n_2}3^{n_3}5^{n_5}$. The pseudo-code 
for a radix-2 recursion of this algorithm looks as shown in algorithm~\ref{doc_dwarf4:alg:fft} (radix-3 and radix-5 
recursions look similar).
\begin{algorithm}[H]
\caption{Factor-2 Cooley-Tukey Fast Fourier Transform algorithm}\label{doc_dwarf4:alg:fft}
\begin{algorithmic}
\vspace{.1cm}
\State \textbf{function} $X_{0,\ldots,n-1}=\textrm{fft}(x_{0,\ldots,n-1})$
\vspace{3mm}
\If{$n=1$}
	\State $X_0=x_0$
\Else
	\State $X_{0,\ldots,n/2-1}=\textrm{fft}(x_{0,2,4,\ldots})$
	\State $X_{n/2,\ldots,n-1}=\textrm{fft}(x_{1,3,5,\ldots})$
	\For{$i=0,\ldots,n/2-1$}
		\State $Y=X_k$
		\State $X_k=Y+\exp(-2\pi i k/n)X_{k+n/2}$
		\State $X_{k+n/2}=Y-\exp(-2\pi i k/n)X_{k+n/2}$
	\EndFor
\EndIf
\end{algorithmic}
\end{algorithm}

\paragraph{Spectral elliptic truncation}
\par
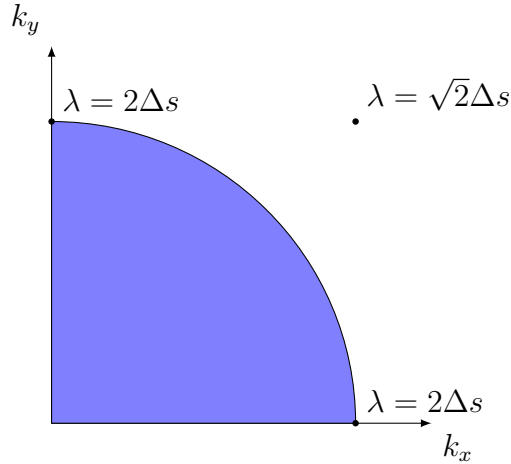
\begin{figure}[htbp]
	\centering
	\begin{tikzpicture}[>=latex]
		\draw[->] (0,0) -- (5,0) node[below right] {$k_x$};
		\draw[->] (0,0) -- (0,5) node[above left] {$k_y$};
		\draw[fill=blue!50] (0,0) -- (4,0) arc[start angle=0, end angle=90, radius=4] -- cycle;
		\draw[fill=black] (0,4) node[above right] {$\lambda=2\Delta s$} circle[radius=1pt];
		\draw[fill=black] (4,0) node[above right] {$\lambda=2\Delta s$} circle[radius=1pt];
		\draw[fill=black] (4,4) node[above right] {$\lambda=\sqrt{2}\Delta s$} circle[radius=1pt];
	\end{tikzpicture}
	\caption{Truncation of the 2D spectrum of a spectral LAM. $\Delta s$ is the grid distance; 
	$k_x$ and $k_y$ are the zonal and meridional wavenumbers, respectively. Without the
	 elliptic truncation, the minimum wavelength $\lambda$ would be $\sqrt{2}\Delta s$ in 
	 the diagonal direction (upper-right point), while it is only $2\Delta s$ in the axes' directions.\label{doc_dwarf4:fig:elliptic_truncation}}
\end{figure}

To ensure an isotropic minimum wavelength, the 2D spectrum of a spectral LAM is truncated 
elliptically, as indicated in figure~\ref{doc_dwarf4:fig:elliptic_truncation}.
\par
Different choices can be made for the maximum wavenumber $k_{max}$ as a function of the 
number of gridpoints $n$. The absolute maximum is the Nyquist wavenumber $k_{max}=\frac{n}{2}$. 
This is called a \emph{linear} grid, where the number of spectral components is equal to the number 
of gridpoints. To limit the effect of aliasing, the number of spectral components can be reduced, 
for instance by taking $k_{max}=\frac{n}{3}$ (\emph{quadratic} grid), or $k_{max}=\frac{n}{4}$ 
(\emph{cubic} grid). Waves with a wavenumber $k>k_{max}$ have zero amplitude.
\paragraph{Periodicity}
\par
Fourier-based spectral methods require that the fields are periodic. This is naturally the case for 
the fields in a global model, but not for the fields in a LAM. Therefore, in a spectral LAM, the fields 
need to be \emph{made} periodic. Following Haugen and Machenhauer \cite{haugen1993}, this 
is achieved by an \emph{extension zone}. This is an artificial extension of the physical domain, 
which is filled in a way that makes the field periodic.
\paragraph{Domain decomposition}
\par
On a distributed-memory machine, the atmospheric fields are distributed over the different 
CPU's (or CPU cores). This means that both the spectral space and the gridpoint space need 
to be decomposed, preferably in a well-balanced way.
\par
Dwarf-D-spectralTransform-BiFourier currently provides the same distribution functionalities 
as the ALADIN LAM. 
\begin{itemize}
	\item in \textbf{spectral space}, the distribution is along the wavenumber (namelist parameter \inlsh{NPRTRW}) and along the different fields (\inlsh{NPRTRV}). The total number of (MPI-)tasks 
	should be \inlsh{NPRTRW}$\times$\inlsh{NPRTRV}.
	\item in \textbf{gridpoint space}, the distribution is along the zonal (\inlsh{NPRGPEW}) and 
	meridional (\inlsh{NPRGPNS}) directions. This means that the complete rectangular domain 
	is decomposed in \inlsh{NPRGPEW}$\times$\inlsh{NPRGPNS} smaller areas, and each area 
	is attributed to one (MPI-)task, as indicated in figure~\ref{doc_dwarf4:fig:domain_decomp}. It should be noted 
	that these areas are not necessarily exactly rectangular. An additional complexity is the treatment 
	of the extension zone, where physics calculations are unnecessary. Therefore, gridpoints in the 
	extension zone should be attributed less weight when determining a well-balanced distribution 
	of the total work load.
	\begin{figure}[htbp]
		\centering
		\input{domain_decomp.tex}
		\caption{Gridpoint distribution of a rectangular LAM domain over 48 MPI tasks.\label{doc_dwarf4:fig:domain_decomp}}
	\end{figure}
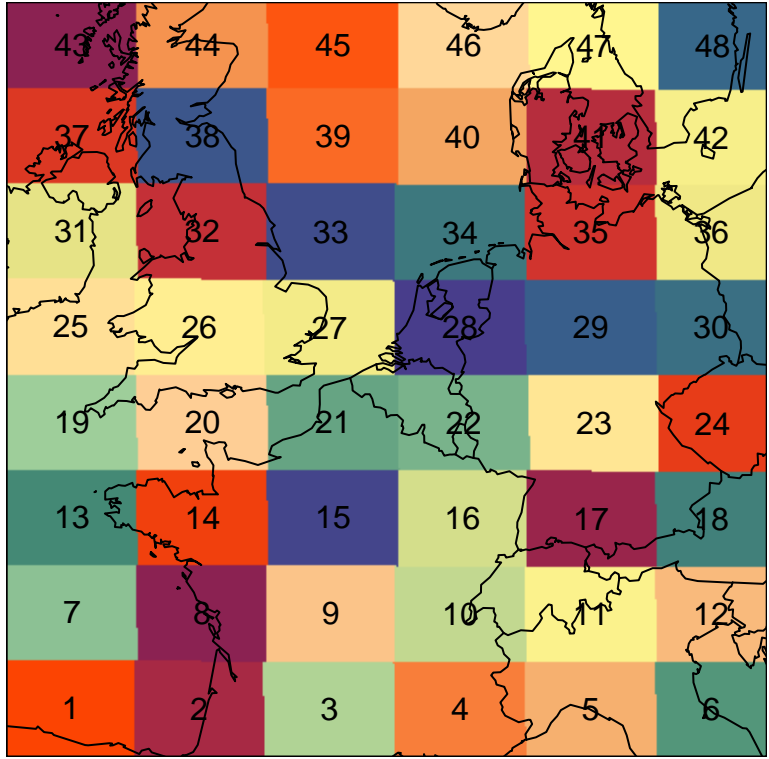
\end{itemize}
\par
In the future, the Atlas framework should be extended with LAM functionalities and take care of this distribution.
\paragraph{I/O interfaces}
\par
The dwarf takes gridpoint data as input. These gridpoint data are read from a grib file. Sample grib files 
with fields of dimensions $200\times180$ and $800\times720$ are available on the ECMWF's ftp server 
(login and password for ESCAPE partners can be found on the Confluence pages). In the longer term, 
larger grids should be envisaged.
\par
The output of this dwarf is the difference (in spectral norm) between the original field, and the field obtained 
after a number of back-and-forth spectral Fourier transforms. This difference is  written to the command-line 
at the end of the program.
\subsubsection{Prototypes}
In this section we describe the prototypes available for this dwarf.

\paragraph{Prototype1}
The first prototype implements the spectral transform required for limited area models without using the Atlas data-structure. The dwarf is suitable for host-type architecture only. 

\subsubsection{Dwarf installation and testing}
\paragraph{Download and installation}
\par
The first step is to download and install the dwarf along 
with all its dependencies. With this purpose, it is possible 
to use the script provided under the ESCAPE software collaboration 
platform:\\
\url{https://git.ecmwf.int/projects/ESCAPE}.

Here you can find a repository called \inlsh{escape}.
You need to download it. There are two options to do this. One option is to use ssh. For this option you need to add an ssh key to your bitbucket account at \url{https://git.ecmwf.int/plugins/servlet/ssh/account/keys}. The link "SSH keys" on this website gives you instructions on how to generate the ssh key and add them to your account. Once this is done you should first create a 
folder named, for instance, ESCAPE, enter into it 
and subsequently download the repository by using the following the steps below:
\begin{lstlisting}[style=BashStyle]
mkdir ESCAPE
cd ESCAPE/
git clone ssh://git@git.ecmwf.int/escape/escape.git
\end{lstlisting}
The other option to download the repo is by using https instead of ssh. Instead of the git command above you then need to use 
\begin{lstlisting}[style=BashStyle]
git clone https://<username>@git.ecmwf.int/scm/escape/escape.git
\end{lstlisting}
where <username> needs to be replace by your bitbucket username.

Once the repository is downloaded into the \inlsh{ESCAPE} folder 
just created, you should find a new folder called \inlsh{escape}. 
The folder contains a sub-folder called \inlsh{bin} that has the 
python/bash script (called \inlsh{escape}) that needs to be 
run for downloading and installing the dwarf and its dependencies. 
To see the various options provided by the script you can type:
\begin{lstlisting}[style=BashStyle]
./escape/bin/escape -h
\end{lstlisting}
To download the dwarf you need to run 
the following command:
\begin{lstlisting}[style=BashStyle]
./escape/bin/escape checkout dwarf-D-spectralTransform-BiFourier \ 
--ssh
\end{lstlisting}
To use https you need to replace --ssh with --user <username>. The commands above automatically check out the \inlsh{develop}
version of the dwarf. If you want to download a specific branch 
of this dwarf, you can do so by typing:
\begin{lstlisting}[style=BashStyle]
./escape/bin/escape checkout dwarf-D-spectralTransform-BiFourier --ssh \
    --version <branch-name>
\end{lstlisting}
An analogous approach can be used for the \inlsh{-{}-user} 
version of the command. You should now have a folder called 
\inlsh{dwarf-D-spectralTransform-BiFourier}.

In the above command, you can specify several other optional 
parameters. To see all these options and how to use them you 
can type the following command:
\begin{lstlisting}[style=BashStyle]
./escape checkout -h
\end{lstlisting}

At this stage it is possible to install the dwarf 
and all its dependencies. This can be done in two 
different ways. The first way is to compile and 
install each dependency and the dwarf separately:
\begin{lstlisting}[style=BashStyle]
./escape/bin/escape generate-install dwarf-D-spectralTransform-BiFourier
\end{lstlisting}
The command above will generate a script 
called \inlsh{install-dwarf-D-spectralTransform-BiFourier} 
that can be run by typing:
\begin{lstlisting}[style=BashStyle]
./install-dwarf-D-spectralTransform-BiFourier
\end{lstlisting}
This last step will build and install the dwarf 
along with all its dependencies in the following 
paths:
\begin{lstlisting}[style=BashStyle]
dwarf-D-spectralTransform-BiFourier/builds/
dwarf-D-spectralTransform-BiFourier/install/
\end{lstlisting}

The second way is to create a bundle that compiles
and installs all the dependencies together:
\begin{lstlisting}[style=BashStyle]
./escape/bin/escape generate-bundle dwarf-D-spectralTransform-BiFourier
\end{lstlisting}
This command will create an infrastructure to avoid
compiling the single third-party libraries individually
when some modifications are applied locally to one of 
them. To complete the compilation and installation process, 
after having run the above command for the bundle, simply 
follow the instructions on the terminal.

In the commands above that generate the installation 
file, you can specify several other optional parameters. 
To see all these options and how to use them you 
can type the following command:
\begin{lstlisting}[style=BashStyle]
./escape generate-install -h
./escape generate-bundle -h
\end{lstlisting}
\paragraph{Testing}
\par
You should now verify that the dwarf works as expected.
With this purpose, we created a testing framework that
allows us to verify that the main features of the dwarf 
are working correctly.
\par
To run this verification, you should run the following 
command:
\begin{lstlisting}[style=BashStyle]
ctest -j<number-of-tasks>
\end{lstlisting}
from inside the \inlsh{builds/dwarf-D-spectralTransform-BiFourier}
folder.
\begin{warningbox}
We strongly advise you to verify via ctest that 
the main functionalities of the dwarf are working 
properly any time you apply modifications to the 
code. Updates that do not pass the tests cannot 
be merged. 
In addition, if you add a new feature to the dwarf,
this should be supported by a test if the existing
testing framework is not already able to verify its
functionality.
\end{warningbox}
For instructions on how to run the executables 
see the next section.
\subsubsection{Run the dwarf}
In this section we provide instructions on how to run the dwarf.
We rename two folders throughout this section as follows:
\begin{lstlisting}[style=BashStyle]
srcs=dwarf-D-spectralTransform-BiFourier/sources/
inst=dwarf-D-spectralTransform-BiFourier/install/
\end{lstlisting}

\paragraph{Executable and namelist file}
\par
The executables of dwarf-D-spectralTransform-BiFourier are found inside
\begin{lstlisting}[style=BashStyle] 
${inst}/dwarf-D-spectralTransform-BiFourier/bin/
\end{lstlisting}
In particular, the executable \inlsh{dwarf-D-spectralTransform-BiFourier-prototype1} needs a namelist file. A sample namelist file can be found under
\begin{lstlisting}[style=BashStyle] 
${srcs}/dwarf-D-spectralTransform-BiFourier/src/tests/fort.4
\end{lstlisting}
\paragraph{Configurable script}
\par
A script that creates a suitable namelist file and launches the executable with appropriate environment settings can be found under
\begin{lstlisting}[style=BashStyle] 
${srcs}/dwarf-D-spectralTransform-BiFourier/src/tests/run_dwarf_D_spectralTransform_BiFourier
\end{lstlisting}
Notable options to this script are
\par
\def\arraystretch{1.5}
\begin{tabular}{@{\hspace*{.05\textwidth}}p{.2\textwidth}@{}p{.75\textwidth}@{}}
	\inlsh{-{}-nproc}	& number of MPI tasks	\\
	\inlsh{-{}-nthread}	&	number of OpenMP threads	\\
	\inlsh{-{}-bin}	&	path to the executable \\[-0.5em]
                        &       \inlsh{dwarf-D-spectralTransform-BiFourier-prototype1}	\\
	\inlsh{-{}-init}	&	path to the initial file; two sample files (on a $180\times200$ grid and on a $720\times800$ grid) are currently present under
\inlsh{\${srcs}/dwarf-D-spectralTransform-BiFourier-testdata/}	\\
	\inlsh{-{}-nfld}	&	number of fields to read from file	\\
	\inlsh{-{}-iters}	&	number of back-and-forth transforms between spectral space and gridpoint space	\\
	\inlsh{-{}-launcher}	&	path to the MPI launcher	\\
\end{tabular}\def\arraystretch{1}
\subsubsection{Integration}
\par
This dwarf provides a stand-alone 2-dimensional spectral Fourier transform, which is an essential building block of spectral limited area models. As the code for this dwarf originates from an actual atmospheric model (IFS/ARPEGE/ALADIN), the integration with atmospheric models is straightforward.
\par
In order to obtain some homogeneity between the different dwarfs defined in the ESCAPE project, they all should be build upon the \Atlas framework. Some work is still necessary in this respect, both on the side of \Atlas (which needs to be extended with LAM functionality), and on the side of this dwarf (which subsequently should be ported to the \Atlas framework). The extension of \Atlas for LAM is deliverable D4.4 of the project.

%% file: domain_decomp.tex
\tikz[x=1cm,y=1cm,inner sep=0pt,outer sep=0pt]{
	\scriptsize
	\draw[use as bounding box] (0,0) ( 10.00000000, 10.00000000);
	\node[above right] at (0,0) {\includegraphics[width= 10.00000000cm,height= 10.00000000cm]{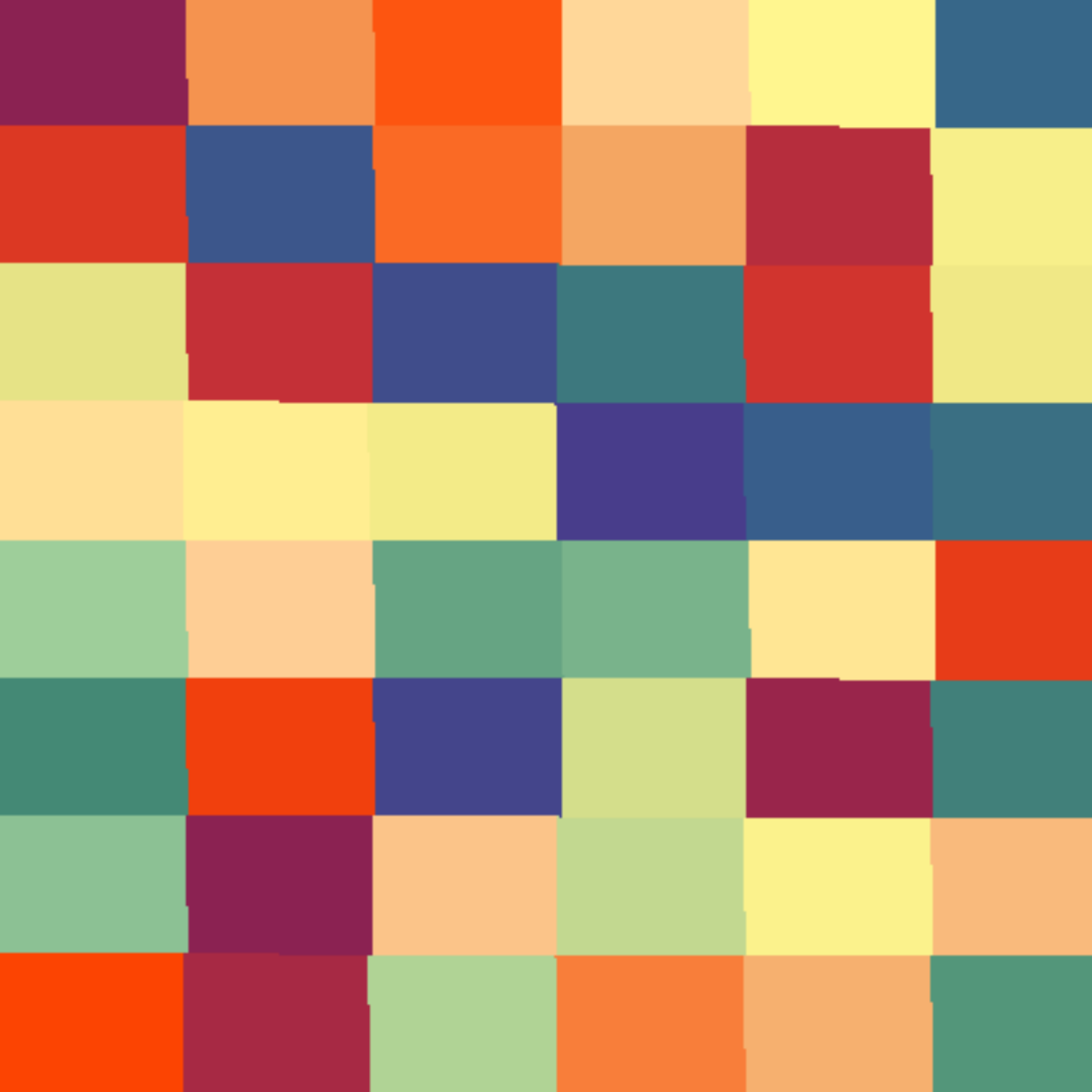}};
	\node[above right] at (0,0) {\includegraphics[viewport= 10.00000000  10.00000000 293.46456693 293.46456693]{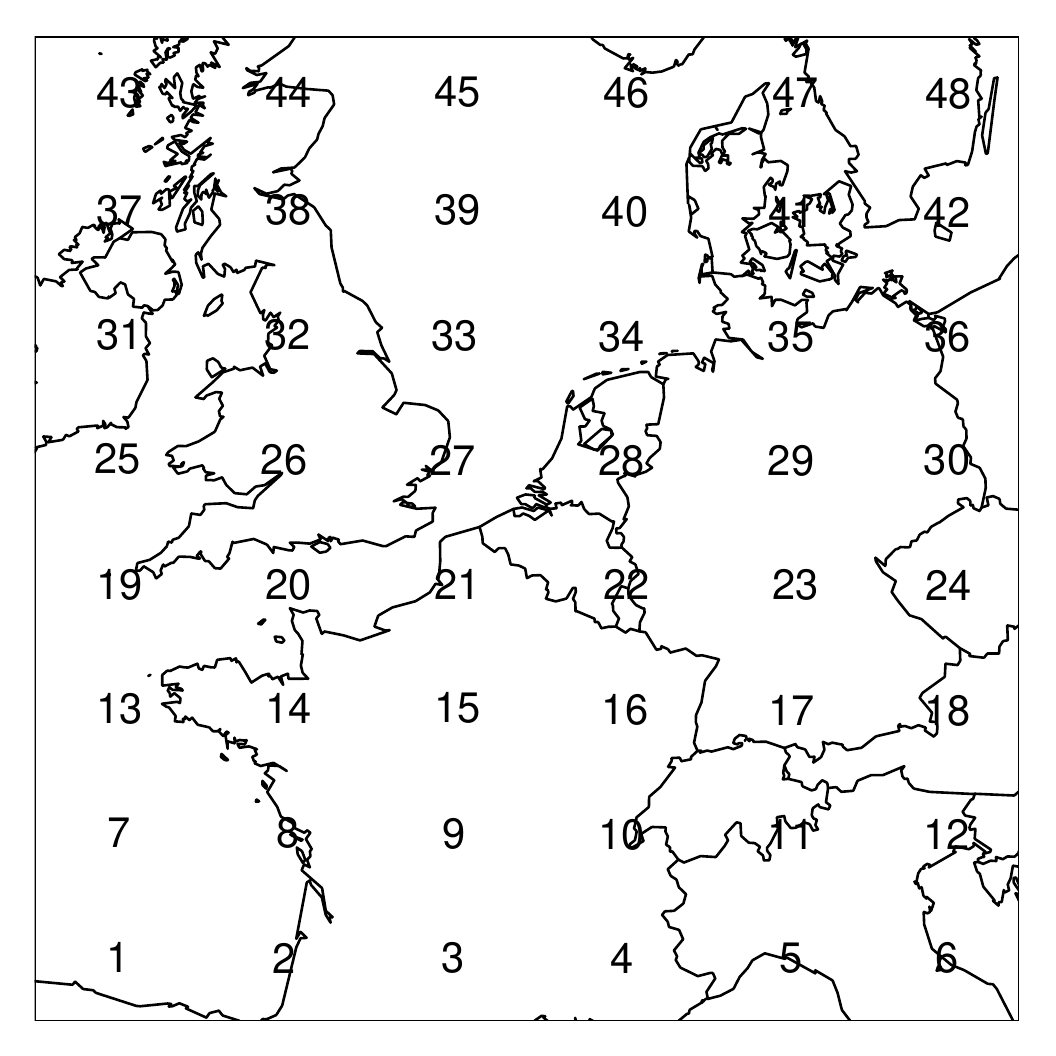}};
}%

%% file: dwarf5.tex
\subsubsection{Scope}
Many NWP and climate models implement Semi-Lagrangian techniques to enable long 
timesteps. Such Semi-Lagrangian schemes in turn need interpolation to determine 
the required quantities. The LAITRI routine provides a 32-point stencil interpolation for Semi-Lagrangian schemes. This code 
is used in IFS and in the limited area model HARMONIE, where it represents approx. 
5\% of runtime and it represents a general infrastructure building block of an NWP 
model.

\subsubsection{Objectives}
The main objectives of this dwarf are to improve the computational time per iteration
of the interpolation step on novel architectures (primarily Xeon Phi and NVIDIA GPUs).

The second objective is to generalize the code to work with new, unstructured grids.
This will take advantage of other work on the Atlas data structure done as part of the 
ESCAPE project, and enable the code to be used in new models, not using the (reduced) 
Gaussian and grid within IFS.

These two aspects are crucial, since a consistent reduction of the computational 
cost might allow these methods, in conjunction with finite-volume discretizations,
to become very competitive for NWP applications, given their compact nature. This 
is particularly attractive for next-generation HPC infrastructures, while retaining 
a semi-implicit time discretization that allows large time-steps.

\subsubsection{Definition of the Dwarf}
Dwarf-I-LAITRI implements LAITRI, a tridimensional 32-point interpolation method
as described in \cite{Ritchie1991}, with optional quasi-monotonic treatment.

Horizontal interpolations are done in both low and high order; the low order 
interpolator is always linear, while the high-order interpolator type is fully controlled 
by a set of weights.
A parameter \inlsh{KQM} determines the monotonicty, with the following values:
\begin{itemize}
\item 0: non-monotonic interpolation
\item 1: horizontally quasi-monotonic interpolation
\item 2: Quasi-monotonic interpolation.
\end{itemize}
%

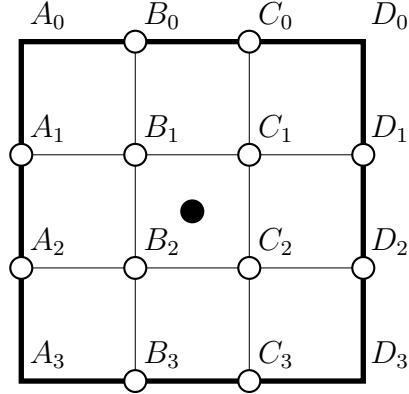
\begin{figure}[H]
\centering
\begin{tikzpicture}[scale=0.5]
\draw [line join=miter, line width=2pt] (0,0) -- (9,0) -- (9,9) -- (0,9) -- cycle {};
\foreach \x in {3,6} {
	\draw [line join=miter, line width=0pt] (\x, 0) -- (\x,9) {};
	\draw [line join=miter, line width=0pt] (0, \x) -- (9,\x) {};
}

\draw[fill=white, thick] (3,0) circle [radius=0.7em];
\draw[fill=white, thick] (6,0) circle [radius=0.7em];
\draw[fill=white, thick] (0,3) circle [radius=0.7em];
\draw[fill=white, thick] (3,3) circle [radius=0.7em];
\draw[fill=white, thick] (6,3) circle [radius=0.7em];
\draw[fill=white, thick] (9,3) circle [radius=0.7em];
\draw[fill=white, thick] (0,6) circle [radius=0.7em];
\draw[fill=white, thick] (3,6) circle [radius=0.7em];
\draw[fill=white, thick] (6,6) circle [radius=0.7em];
\draw[fill=white, thick] (9,6) circle [radius=0.7em];
\draw[fill=white, thick] (3,9) circle [radius=0.7em];
\draw[fill=white, thick] (6,9) circle [radius=0.7em];
\draw[fill=black, thick] (4.5,4.5) circle [radius=0.7em];

\node at (0.7,0.7) { $A_3$};
\node at (3.7,0.7) { $B_3$};
\node at (6.7,0.7) { $C_3$};
\node at (9.7,0.7) { $D_3$};
\node at (0.7,3.7) { $A_2$};
\node at (3.7,3.7) { $B_2$};
\node at (6.7,3.7) { $C_2$};
\node at (9.7,3.7) { $D_2$};
\node at (0.7,6.7) { $A_1$};
\node at (3.7,6.7) { $B_1$};
\node at (6.7,6.7) { $C_1$};
\node at (9.7,6.7) { $D_1$};
\node at (0.7,9.7) { $A_0$};
\node at (3.7,9.7) { $B_0$};
\node at (6.7,9.7) { $C_0$};
\node at (9.7,9.7) { $D_0$};
\end{tikzpicture}
\vspace{5mm}
\caption{12-Point Horizontal Interpolation Stencil}
\label{doc_dwarf5:12point}
\end{figure}

In fact, the array PDLO contains the precomputed linear interpolation weights 
corresponding to the central longitude for a given western point.  These weights 
are referred to as `zonal weights' of a given number (western point).  For example, 
for western point 1 of the first linear interpolation at stencil level 0, the value of 
PDLO(jrof,jlev,1) will be
\[
PDLO(jrof,jlev,1) = ZDLO1 = \frac{\Lambda_O-\Lambda_{B_1}}{\Lambda_{C_1}-\Lambda_{B_1}}.
\]
where $\Lambda$ is a longitude on the computational sphere corresponding to a particular 
point in the stencil.\\

The functions $f_2(\alpha)$, $f_3(\alpha)$ and $f_4(\alpha)$ correspond to the Lagrange 
basis polynomials in the variable $\alpha$, which itself is dependent on the position of the 
interpolation point. This position is determined before LAITRI is called, and so $f_2(\alpha)$ 
etc. are precomputed, the values residing in the array PCLO. In fact, for interpolation of points 
$A_1$, $B_1$, $C_1$, $D_1$, the value of $\alpha$ is computed as
\[
\alpha = ZDLO1 = \frac{\Lambda_O-\Lambda_{B_1}}{\Lambda_{C_1}-\Lambda_{B_1}}
\]

\noindent In order that we may fix basis polynomials, we stipulate that the horizontal positions 
(values of $\alpha$) of $A_1$, $B_1$, $C_1$, $D_1$ are $-1$, $0$, $1$, $2$, respectively. 
Then, our basis polynomials (recalling that we have fixed the first at unity) are
\begin{align*}
f_2(\alpha) &= PCLO(jrof,jlev,1,1) = \cfrac{(\alpha + 1)(\alpha - 2)(\alpha - 1)}{2}\\
f_3(\alpha) &= PCLO(jrof,jlev,2,1) = \cfrac{-(\alpha + 1)(\alpha - 2)\alpha}{2}\\
f_4(\alpha) &= PCLO(jrof,jlev,3,1) = \cfrac{\alpha(\alpha - 1)(\alpha + 1)}{6}
\end{align*}

Linear interpolation weights PDLAT and cubic interpolation weights PCLA are precomputed. 
Finally, the vertical cubic interpolation weights PVINTW are also precomputed.

\paragraph{Pseudo-algorithm}
%
%
\begin{algorithm}[H]
\caption{Core LAITRI 32 points 3D interpolation routine}\label{doc_dwarf5:alg:core-loop}
\begin{algorithmic}
\begin{footnotesize}
\For{jlev = 1,...,kflev}
\For{jrof = kst,...,kprof}
\vspace{.2cm}
\LineComment{zonal interpolation at stencil level 0:}
\Indent
\vspace{-.2cm}
\State ---------------------------------------------------------------------------------------
\vspace{0cm}
\State $\displaystyle 
Z_{10} = X_{B_1} + \frac{\Lambda_O-\Lambda_{B_1}}{\Lambda_{C_1}-\Lambda_{B_1}} \left(X_{C_1} - X_{B_1}\right)$
\vspace{.2cm}
\State $\displaystyle
Z_{20} = X_{B_2} + \frac{\Lambda_O-\Lambda_{B_2}}{\Lambda_{C_2}-\Lambda_{B_2}} \left(X_{C_2} - X_{B_2}\right)$
\vspace{0.1cm}
\State ---------------------------------------------------------------------------------------
\vspace{.4cm}
\EndIndent
\State ...
\algstore{laitriAlg}
\end{footnotesize}
\end{algorithmic}
\end{algorithm}

\begin{algorithm}[H]
\begin{algorithmic}
\begin{footnotesize}
\algrestore{laitriAlg}
\vspace{.2cm}
\State ...
\LineComment{zonal interpolation at stencil level 1:}
\Indent
\vspace{-.2cm}
\State ---------------------------------------------------------------------------------------
\vspace{0cm}
\State $\displaystyle 
Z_{01} = X_{B_0} + \frac{\Lambda_O-\Lambda_{B_0}}{\Lambda_{C_0}-\Lambda_{B_0}} \left(X_{C_0} - X_{B_0}\right)$
\vspace{.2cm}
\State $\displaystyle 
Z_{11} = X_{A_1} + f_2(\alpha)(X_{B_1} - X_{A_1}) + f_3(\alpha)(X_{C_1} - X_{A_1}) + f_4(\alpha)(X_{D_1} - X_{A_1})$
\vspace{.2cm}
\State $\displaystyle 
Z_{21} = X_{A_1} + f_2(\alpha)(X_{B_1} - X_{A_1}) + f_3(\alpha)(X_{C_1} - X_{A_1}) + f_4(\alpha)(X_{D_1} - X_{A_1})$
\vspace{.2cm}
\State $\displaystyle 
Z_{31} = X_{B_3} + \frac{\Lambda_O-\Lambda_{B_3}}{\Lambda_{C_3}-\Lambda_{B_3}} \left(X_{C_3} - X_{B_3}\right)$
\vspace{0.1cm}
\State ---------------------------------------------------------------------------------------
\vspace{.4cm}
\EndIndent
\LineComment{zonal interpolation at stencil level 2:}
\Indent
\vspace{-.2cm}
\State ---------------------------------------------------------------------------------------
\vspace{0cm}
\State $\displaystyle 
Z_{02} = X_{B_0} + \frac{\Lambda_O-\Lambda_{B_0}}{\Lambda_{C_0}-\Lambda_{B_0}} \left(X_{C_0} - X_{B_0}\right)$
\vspace{.2cm}
\State $\displaystyle 
Z_{12} = X_{A_1} + f_2(\alpha)(X_{B_1} - X_{A_1}) + f_3(\alpha)(X_{C_1} - X_{A_1}) + f_4(\alpha)(X_{D_1} - X_{A_1})$
\vspace{.2cm}
\State $\displaystyle 
Z_{22} = X_{A_1} + f_2(\alpha)(X_{B_1} - X_{A_1}) + f_3(\alpha)(X_{C_1} - X_{A_1}) + f_4(\alpha)(X_{D_1} - X_{A_1})$
\vspace{.2cm}
\State $\displaystyle 
Z_{32} = X_{B_3} + \frac{\Lambda_O-\Lambda_{B_3}}{\Lambda_{C_3}-\Lambda_{B_3}} \left(X_{C_3} - X_{B_3}\right)$
\vspace{0.1cm}
\State ---------------------------------------------------------------------------------------
\vspace{.4cm}
\EndIndent
\LineComment{zonal interpolation at stencil level 3:}
\Indent
\vspace{-.2cm}
\State ---------------------------------------------------------------------------------------
\vspace{0cm}
\State $\displaystyle 
Z_{13} = X_{B_1} + \frac{\Lambda_O-\Lambda_{B_1}}{\Lambda_{C_1}-\Lambda_{B_1}} \left(X_{C_1} - X_{B_1}\right)$
\vspace{.2cm}
\State $\displaystyle
Z_{23} = X_{B_2} + \frac{\Lambda_O-\Lambda_{B_2}}{\Lambda_{C_2}-\Lambda_{B_2}} \left(X_{C_2} - X_{B_2}\right)$
\vspace{0.1cm}
\State ---------------------------------------------------------------------------------------
\vspace{.4cm}
\EndIndent
\LineComment{meridional interpolation at the four stencil levels:}
\Indent
\vspace{-.2cm}
\State ---------------------------------------------------------------------------------------
\vspace{0cm}
\State $\displaystyle 
Z_0 = Z_{10} + PDLAT(jrof,jlev) \left(Z_{20} - Z_{10}\right)$
\vspace{.2cm}
\State $\displaystyle 
Z_1  = Z_{01} + PCLA(jrof,jlev,1)\left(Z_{11} - Z_{01}\right) $ 
\State $++ PCLA(jrof,jlev,2)\left(Z_{21} - Z_{01}\right) + PCLA(jrof,jlev,3)\left(Z_{31} - Z_{01}\right)$
\vspace{.2cm}
\State $\displaystyle 
Z_2  = Z_{02} + PCLA(jrof,jlev,1)\left(Z_{12} - Z_{02}\right) + $
\State $+ PCLA(jrof,jlev,2)\left(Z_{22} - Z_{02}\right) + PCLA(jrof,jlev,3)\left(Z_{32} - Z_{02}\right)$
\vspace{.2cm}
\State $\displaystyle 
Z_3 = Z_{13} + PDLAT(jrof,jlev) \left(Z_{23} - Z_{13}\right)$
\vspace{0.1cm}
\State ---------------------------------------------------------------------------------------
\vspace{.4cm}
\EndIndent
\LineComment{interpolation along the vertical direction:}
\Indent
\vspace{-.2cm}
\State ---------------------------------------------------------------------------------------
\vspace{0cm}
\State $PXF(jrof,jlev) = Z_0 + PVINTW(jrof,jlev,1)(Z_1 - Z_0) +$
\State $ + PVINTW(jrof,jlev,2)(Z_2 - Z_0) + PVINTW(jrof,jlev,3)(Z_3 - Z_0)$
\vspace{0.1cm}
\State ---------------------------------------------------------------------------------------
\vspace{.0cm}
\EndIndent
\EndFor
\EndFor
\end{footnotesize}
\end{algorithmic}
\end{algorithm}

\paragraph{I/O interfaces}

The input/output (I/O) interfaces for this dwarf are as follows.
\begin{itemize}
\item \textbf{Input}: variable (field) to be interpolated. 
\item \textbf{Output}: Interpolated variable.
\end{itemize}

\subsubsection{Prototypes}
In this secion we describe the prototypes available for this dwarf.

\paragraph{Prototype 1}
The first prototype implements the dwarf using the Atlas data-structure.  Currently, this 
is work-in-progress as it is unclear at present whether we can guarantee that certain 
data structures will be provided so as to make certain optimizations for target platforms 
such as Xeon Phi/Knights Landing possible.  It may be the case that Laitri itself will not 
need to implement Atlas functionality, for example if the Atlas grid values are computed 
prior to calling Laitri, and then merely passed in to Laitri in a pre-populated array.

\paragraph{Prototype 2}
The second prototype implements the dwarf without using the Atlas data-structure. 
This was the logical starting-point for performance experiments as this required only 
minimal modification to the Laitri program.  We have summarized the results of our 
experiments in a separate document referring to Prototype 2.

\subsubsection{Dwarf usage and testing}
In this section we describe how to download and install 
the dwarf along with all its dependencies and we show 
how to run it for a simple test case.

\paragraph{Download and installation}
The first step is to download and install the dwarf along 
with all its dependencies. With this purpose, it is possible 
to use the script provided under the ESCAPE software collaboration 
platform:\\
\url{https://git.ecmwf.int/projects/ESCAPE}.

Here you can find a repository called \inlsh{escape}.
You need to download it. There are two options to do this. One option is to use ssh. For this option you need to add an ssh key to your bitbucket account at \url{https://git.ecmwf.int/plugins/servlet/ssh/account/keys}. The link "SSH keys" on this website gives you instructions on how to generate the ssh key and add them to your account. Once this is done you should first create a 
folder named, for instance, ESCAPE, enter into it 
and subsequently download the repository by using the following the steps below:
\begin{lstlisting}[style=BashStyle]
mkdir ESCAPE
cd ESCAPE/
git clone ssh://git@git.ecmwf.int/escape/escape.git
\end{lstlisting}
The other option to download the repo is by using https instead of ssh. Instead of the git command above you then need to use 
\begin{lstlisting}[style=BashStyle]
git clone https://<username>@git.ecmwf.int/scm/escape/escape.git
\end{lstlisting}
where <username> needs to be replace by your bitbucket username.

Once the repository is downloaded into the \inlsh{ESCAPE} folder 
just created, you should find a new folder called \inlsh{escape}. 
The folder contains a sub-folder called \inlsh{bin} that has the 
python/bash script (called \inlsh{escape}) that needs to be 
run for downloading and installing the dwarf and its dependencies. 
To see the various options provided by the script you can type:
\begin{lstlisting}[style=BashStyle]
./escape/bin/escape -h
\end{lstlisting}
To download the dwarf you need to run 
the following command:
\begin{lstlisting}[style=BashStyle]
./escape/bin/escape checkout dwarf-I-LAITRI \ 
--ssh
\end{lstlisting}
To use https you need to replace --ssh with --user <username>. The commands above automatically check out the \inlsh{develop}
version of the dwarf. If you want to download a specific branch 
of this dwarf, you can do so by typing:
\begin{lstlisting}[style=BashStyle]
./escape/bin/escape checkout dwarf-I-LAITRI --ssh \
--version <branch-name>
\end{lstlisting}
Specific versions of the dwarf code are being developed for
Xeon Phi and NVidia GPUs. Currently a branch exists for the MIC
architecture; this is identical to the \inlsh{develop} branch
except for being called \inlsh{feature/mic}. Checking this out as above
builds a version for the Xeon Phi (tested on Knights Landing).
Analogous approach can be used for the \inlsh{-\,-user} 
version of the command. You should now have a folder called 
\inlsh{dwarf-I-LAITRI}.

In the above command, you can specify several other optional 
parameters. To see all these options and how to use them you 
can type the following command:
\begin{lstlisting}[style=BashStyle]
./escape checkout -h
\end{lstlisting}

At this stage it is possible to install the dwarf 
and all its dependencies. This can be done in two 
different ways. The first way is to compile and 
install each dependency and the dwarf separately:
\begin{lstlisting}[style=BashStyle]
./escape/bin/escape generate_install dwarf-I-LAITRI
\end{lstlisting}
The command above will generate a script 
called \inlsh{install-dwarf-I-LAITRI} 
that can be run by typing:
\begin{lstlisting}[style=BashStyle]
./install-dwarf-I-LAITRI
\end{lstlisting}
This last step will build and install the dwarf 
along with all its dependencies in the following 
paths:
\begin{lstlisting}[style=BashStyle]
dwarf-I-LAITRI/builds/
dwarf-I-LAITRI/install/
\end{lstlisting}

The second way is to create a bundle that compile 
and install all the dependencies together:
\begin{lstlisting}[style=BashStyle]
./escape/bin/escape generate_bundle dwarf-I-LAITRI
\end{lstlisting}
This command will create an infrastructure to avoid
compiling the single third-party libraries individually
when some modifications are applied locally to one of 
them. To complete the compilation and installation process, 
after having run the above command for the bundle, simply 
follow the instructions on the terminal.

In the commands above that generate the installation 
file, you can specify several other optional parameters. 
To see all these options and how to use them you 
can type the following command:
\begin{lstlisting}[style=BashStyle]
./escape generate-install -h
./escape generate-bundle -h
\end{lstlisting}

\paragraph{Testing}
You should now verify that the dwarf works as expected.
With this purpose, we created a testing framework that
allows us to verify that the main features of the dwarf 
are working correctly.

To run this verification, you should run the following 
command:
\begin{lstlisting}[style=BashStyle]
ctest -j<number-of-tasks>
\end{lstlisting}
from inside the \inlsh{builds/dwarf-I-LAITRI}
folder.
\begin{warningbox}
We strongly advise you to verify via ctest that 
the main functionalities of the dwarf are working 
properly any time you apply modifications to the 
code. Updates that do not pass the tests cannot 
be merged. 
In addition, if you add a new feature to the dwarf,
this should be supported by a test if the existing
testing framework is not already able to verify 
its functionality.
\end{warningbox}
For instructions on how to run the executables 
see the next section.

\subsubsection{Run the dwarf}
To run the dwarf, driver programs \inlsh{dwarf-I-LAITRI-laitri} 
and \inlsh{laitri-atlas} are provided, which are 
automatically created as the program is built. These will take test parameters via namelist
on the command-line, and the source code \inlsh{driver.F90} shows how to all the laitri library.

\paragraph{A note on bit reproducibility}
The LAITRI Dwarf (like the forecast model) has been written to always produce 
{\bfseries BIT IDENTICAL RESULTS} when either the number of MPI tasks 
is changed or the number of OpenMP threads is changed provided suitable 
compilation options and blas library have been used.
As a minimum benchmarkers are required to only use compiler options such 
that if a job is re-run with the same executable and with no change in either 
input data or namelist parameters, it must give results which are bit-identical 
with those of the first run.

This benchmark has been built and tested using Cray, Intel and GNU compilers. 
Each build satisfied both the correctness and bit reproducibility criteria. Results 
were fully reproducible for all tested combinations of tasks and threads. The 
optimisation options used were

Cray  - {\bfseries -O thread1 -hflex\_mp=conservative -hfp1 -hadd\_paren}\\
Intel - {\bfseries -O2 -fp-model precise -fp-speculation=strict}\\
GNU   - {\bfseries -O2}\\

ECMWF recognise that the requirement to achieve full bit reproducibility when the 
number of threads or tasks change will require conservative compilation options 
with an impact of performance. Benchmarkers are free to chose more aggressive 
options provided the minimum requirement of identical results on a re-run is still 
achieved.
\\

\subsubsection{Integration}

Given that the host version of the LAITRI dwarf is derived from the IFS source (code
cycle CY41R1) there should be a minimal effort to integrate any modifications into 
an official IFS source cycle.

On the other hand, modifications for a GPU device or Xeon Phi implementation 
(in offload mode)  of the dwarf are only meaningful if the modifications are consistent 
with a more complete IFS port to a GPU device. Programming productivity should 
be considered for such a port taking into account features such as deep copy 
of derived type objects.
The dwarf should always be run with multiple tasks and over multiple nodes, 
to assess the effect of any potential overlap of computations with MPI communications.

Of concern for a GPU device implementation is the portability of a future IFS model 
where OpenACC parallelisation directives are applied to every DO loop nest, or 
handled by a DSL approach if that is pursued.

%% file: dwarf6.tex
\subsubsection{Scope}

A key component of any NWP dynamical core is the advection
scheme. Its purpose is to solve the PDEs modelling the transport of
momentum, heat and mass on a spherical
domain. The semi-Lagrangian (SL) method is a very efficient technique
for solving such transport equations mainly
because of its unconditional stability and good dispersion properties
which  permit accurate integrations using long timesteps. However, it
is known that due to communication overheads the efficiency of the SL
method reduces as resolution increases towards cloud resolving scales 
and computer architectures move towards exascale platforms. 

Dwarf-D-advection-SemiLagrangian implements a stand-alone (completely 
autonomous from a full NWP model) SL advection scheme for the passive 
(without forcing terms) advection of tracer fields on a 3D spherical domain 
without orography using a prescribed wind field.
Given that developing and testing new ideas inside a complex and mature 
NWP system is a lengthy process, the dwarf provides a simplified environment 
for (i) testing the computational performance and scalability of the SL method 
on different super-computing hardware platforms; for (ii) exploring various 
implementation aspects of the SL technique and for (iii) comparing it against 
other established advection methods such as Finite Volume.

\subsubsection{Objectives}

The main objective of this dwarf is to assess the scalability limits of the SL technique 
isolating it from other NWP model components, such as the semi-implicit solver and 
parametrizations.

Development and evaluation of dwarf-D-advection-SemiLagrangian will include:
\begin{enumerate}
\item Developing a solver for the passive 3D advection problem on an spherical 
coordinate set up without orography using a prescribed wind field. The spherical 
domain is discretized using a quasi-uniform reduced Gaussian grids. Regular 
lat-lon grids (latitude shifted) would be also available for testing.
\item Developing a Fortran 2003 code using the NWP software framework Atlas.
\item Developing the capability to run on multiple hosts (e.g. CPUs) of devices 
(e.g. GPUs) using MPI and multiple threads.
\item Test the SL approach on different hardware platforms using standard 
test cases (e.g. ``solid body rotation'' and ``Rossby-Haurwitz'' wave test) in 
order to identify the best solutions in terms of computational time per timestep.
Among the best solutions, it will also be necessary to identify the best compromise 
in terms of energy cost. 
\end{enumerate}

\subsubsection{Definition of the Dwarf}
Dwarf-D-advection-SemiLagrangian implements a three-dimensional semi-Lagrangian 
advection scheme. It solves the following transport equation in Lagrangian form:
\begin{equation}
\frac{D\phi_{\chi}}{Dt} = 0, \qquad
\displaystyle{\frac{D}{Dt}=\frac{\partial}{\partial t} + 
  \mathbf{V}\cdot \nabla}, \;\; \mathbf{V}=(u,v,w)
\label{doc_dwarf6:eq:sladv} 
\end{equation}
where $\phi_{\chi}$ is a scalar field advected by the wind
$\mathbf{V}=(u,v,w)$. If the field is a tracer then $\phi_{\chi}$ should represent its mixing 
or specific ratio i.e. $\phi_{\chi}=\rho_{\chi}/\rho_{d}$ or $\phi_{\chi}=\rho_{\chi}/\rho$ 
where $\rho$, $\rho_d$ the density of air and dry air respectively and $\rho_{\chi}$ 
the density of the tracer. Solving Eq. \eqref{doc_dwarf6:eq:sladv} then is equivalent to solving 
the continuity equation for the tracer:
\begin{equation}
\frac{D\rho_{\chi}}{Dt} = -\rho_{\chi}\nabla\cdot\mathbf{V}. \label{doc_dwarf6:eq:sladv2} 
\end{equation}
The advection problem  \eqref{doc_dwarf6:eq:sladv} is formulated in spherical coordinates and discretized 
in a quasi-uniform (Gaussian) or uniform grid. For the dwarf, it is assumed that the terrain 
is ``flat'' i.e. there is no orography and that the wind field is prescribed. In what follows we 
briefly outline the SL scheme.

To solve Eq. \eqref{doc_dwarf6:eq:sladv} we integrate along the trajectory of a fluid parcel in the time 
interval $[t,t+\Delta t]$ \[ \int_t^{t+\Delta t} D\phi_{\chi} = 0, \] to obtain:
\begin{equation}
 \phi_{\chi,a}^{t+\Delta t}=\phi_{\chi,d}^{t}. \label{doc_dwarf6:eq:sol}
\end{equation}
Subscript letters $a$, $d$ denote the so-called arrival and departure
points. The former is the location of a parcel at time $t+\Delta t$
and coincides with a grid-point and the latter its location at time
$t$. In the equations that follow, for notational convenience, the arrival
point subscript $a$ will be omitted. The departure point (d.p.) lies
somewhere in the space between grid-points and has to be
found. Eq. \eqref{doc_dwarf6:eq:sol} implies that the solution at next timestep
is the value of the available (at time $t$) field at the
d.p. Therefore, a semi-Lagrangian scheme performs the
following steps:
\begin{enumerate}
\item For each grid-point solve the trajectory equation to determine the
  departure point (d.p.) $\mathbf{r_d}=(\lambda_d,\theta_d,\eta_d)$:
\begin{equation}
\frac{D\mathbf{r}}{Dt}=\mathbf{V}(\mathbf{r},t) \Rightarrow 
\underbrace{\mathbf{r}}_{arrival\;g.p.}-\underbrace{\mathbf{r}_d}_{unknown\;d.p.}
=\int_t^{t+\Delta t} \mathbf{V}(\mathbf{r},t)dt \label{doc_dwarf6:eq:dp1}
\end{equation} where $\lambda$, $\theta$, $\eta$ is longitude,
latitude and the vertical coordinate used respectively. 
\item Interpolate $\phi_{\chi}$ to $\mathbf{r_d}$ to obtain
\begin{equation} 
\phi^{t+\Delta t}_{\chi} = \phi^t_{\chi,d} \label{doc_dwarf6:eq:interp}
\end{equation}
\end{enumerate}

\paragraph{Finding departure points}

Solving the trajectory equation \eqref{doc_dwarf6:eq:dp1} requires the
numerical approximation of a velocity integral. The mid-point rule is
a commonly used approach:

\begin{equation}
\mathbf{r}-\mathbf{r_d} = \Delta t \:
\mathbf{V}\left(\frac{\mathbf{r+r_d}}{2},t+\frac{\Delta t}{2}
\right). \label{doc_dwarf6:midp}
\end{equation}

The time-dependent discrete trajectory equation
\eqref{doc_dwarf6:midp} must be solved for the d.p. $\mathbf{r_d}$. The velocity
field at the
trajectory mid-point and at time $t+\Delta t/2$ must be found first. 
A usual practice is to extrapolate in time the velocity and then interpolate the
derived field (with linear interpolation) at the estimated mid-point. The second
order time-extrapolation formula is often used:
\[ \mathbf{V}^{t+\Delta t/2} = 1.5 \mathbf{V}^t - 0.5 \mathbf{V}^{t-\Delta t} + O(\Delta t^2) \]
while the fixed-point iteration Algorithm
\ref{doc_dwarf6:alg:midpoint} can be used to compute the departure point.
\begin{algorithm}
\caption{Departure point calculation based on mid-point iterative
  scheme (for simplicity a Cartesian coordinate framework description
  is used).}\label{doc_dwarf6:alg:settls}
\begin{algorithmic}
\State{Extrapolate and store: $\mathbf{V}^{t+\Delta t/2} = 1.5 \mathbf{V}^t - 0.5\mathbf{V}^{t-\Delta t}$}
\State{Initialise: $\mathbf{r_d}^{(1)}=\mathbf{r}-\Delta t \mathbf{V}^t$}
\For{$\nu=2,\ldots,\nu_{max}$}
\State{Interpolate $\mathbf{V}^{t+\Delta t/2}$ to midpoint: \newline
  \hspace*{8mm} $\mathbf{V}^{t+\Delta t/2}_m\equiv \mathbf{V}^{t+\Delta
    t/2}(\mathbf{r_m}^{(\nu-1)}), \quad \mathbf{r_m}^{(\nu-1)}\equiv 0.5[\mathbf{r+r_d^{(\nu-1)}}]$}
\State{Update: \( \mathbf{r_d}^{(\nu)}=\mathbf{r}-\Delta t
  \mathbf{V}^{t+\Delta t/2}_m \)}
\EndFor
\end{algorithmic} \label{doc_dwarf6:alg:midpoint}
\end{algorithm}

The ECMWF model IFS, employs a
scheme based on SETTLS extrapolation formula (Stable Two-Time-Level
Extrapolation Scheme) \cite{Hortal2002}:
\[ \mathbf{r_d} = \mathbf{r} - \frac{\Delta t}{2} \left(\mathbf{V}^t
  +\left[2\mathbf{V}^t-\mathbf{V}^{t-\Delta t} \right]_d \right) \]
This is also solved iteratively as described by Algorithm \ref{doc_dwarf6:alg:settls}.
\begin{algorithm}
\caption{Departure point calculation based on SETTLS (for simplicity
  a Cartesian coordinate framework description is used).}
\begin{algorithmic}
\State{Compute and store: $\mathbf{V}^{*}=2\mathbf{V}^t-\mathbf{V}^{t-\Delta t}$}
\State{Initialise: $\mathbf{r_d}^{(1)}=\mathbf{r}-\Delta t \mathbf{V}^t$}
\For{$\nu=2,\ldots,\nu_{max}$}
\State{Interpolate $\mathbf{V}^{*}$ to the d.p.:
  $\mathbf{V_d^{*}}^{(\nu-1)}\equiv V^{*}(\mathbf{r_d}^{(\nu-1)})$ }
\State{Update: \( \mathbf{r_d}^{(\nu)}=\mathbf{r}-0.5\Delta t
  \left( \mathbf{V^t} +\mathbf{V_d^{*}}^{(\nu-1)}\right) \)}
\EndFor
\end{algorithmic} 
\end{algorithm}

A sufficient condition 
for convergence of the iterative procedure of Algorithms
\ref{doc_dwarf6:alg:midpoint}, \ref{doc_dwarf6:alg:settls} is given in
\cite{PBS1985}. This is a Lipschitz condition and requires that
the timestep $\Delta t$ should be smaller than the 
reciprocal of the absolute maximum value of the wind-shear at each
direction (see \cite{SC1991}). The geometric interpretation of this
condition (see \cite{SP1992}) is that trajectories do not intersect
each-other. In practice this is satisfied for atmospheric flows. We
need at least two iterations to obtain second order accuracy. For
flows with high CFL numbers, strong wind shear and large timesteps
further iterations may be necessary for convergence.

Both the mid-point and SETTLS iterative method have been
implemented for the semi-Lagrangian dwarf. The 
2-iteration version of the former is equivalent to the 2nd order
Runge-Kutta scheme RK2 described in chapter 6 of  
\cite{Durran99}. Calculating the d.p. in a spherical domain adds
further complexity in the algorithm. This can be handled by different
methods, for example: (i) solving the d.p. in a rectangular $(\lambda, \theta)$
domain (ii) using Ritchie's transformed cartesian coordinate
method \cite{R1987} or (iii) using the ``rotation matrix'' method (see appendix of
\cite{THS2001}).  The first one is inaccurate away from the equator
and was only available at the initial release 0.1.0 for testing
purposes. Methods (ii) and (iii) are available in the current version with the latter being
the default as it is the method used by the ECMWF model.

\paragraph{Interpolation}
Essentially, through the SL discretization the
advection problem is turned to an interpolation one. 
Although linear interpolation is sufficiently accurate for the wind
component interpolations needed when computing the d.p., a high order
scheme must be used for interpolating the transported field $\phi$ at
the d.p. (see Eq. \eqref{doc_dwarf6:eq:interp}). As noted by \cite{SC1991} cubic
Lagrange has been found to be a good compromise between computational
cost and accuracy. 

To interpolate to the d.p. we need to define first an interpolation
``stencil'' i.e. the set of points surrounding the d.p. that the
interpolation formula will use. The stencil is always centred at the
departure point. We currently have available three options for the dwarf: (i) an 8-point stencil tri-linear interpolation scheme; (ii) a
64-point stencil tri-cubic-Lagrange interpolation scheme and (iii) the
32-point 
stencil ECMWF quasi tri-cubic-Lagrange scheme. Interpolation to a d.p. 
$(\lambda_d,\theta_d,\eta_d)$ is done as a sequence of three separate
one-dimensional interpolations in $\lambda_d$, $\theta_d$ and
$\eta_d$. The aim of the first interpolation is to find the field
values at the d.p. longitude. These interpolated values,
all at the same longitude $\lambda_d$, are then
interpolated to the d.p. latitude $\theta_d$. Finally, the
outcome of the previous interpolations which represents values at
the same longitude and latitude (vertically aligned) is interpolated
to $\eta_d$ to obtain the interpolated field value at the three
dimensional departure point. For the
ECMWF quasi-cubic scheme, linear interpolation is
used at the two external latitude rows of the stencil of the two
adjacent levels above and below the d.p. while cubic for the
remaining (interior) rows. Cubic interpolation is used in latitude and
in the vertical. For the top and bottom levels linear interpolation is
used in longitude and latitude. Overall, this is cheaper than standard
tri-cubic Lagrange and has been found to have similar
accuracy. 

Finally, it is possible to use a quasi-monotone limiter such as the one
by \cite{BS1992} to avoid generating new maxima or minima in the
solution and avoid unphysical oscillations (shape preservation).

\paragraph{Parallelization}

The spherical domain is decomposed to a set of almost equal area
sub-domains with an overlapping halo region. Each sub-domain is
assigned to an MPI task where multiple OpenMP threads are used to
achieve a further shared memory layer of parallelization. The
currently implemented MPI parallelization strategy is based on a
``max wind halo'' approach in which a halo sufficiently large to
perform the SL calculations is used without need to exchange data between
MPI tasks. Two other methods that can be coded and tested:
\begin{itemize}
\item `Insourcing'. This is a communication on demand approach
  currently used by the SL avection scheme of IFS in combination with
  max-wind halo. In this strategy when the d.p. lies outside the
  domain of a local MPI task then the remote task holding 
  the required data is instructed to transmit these to the local and
  the latter completes the calculation. 
\item `Outsourcing'. This is an alternative communication on demand
  strategy. The local MPI task instructs the remote one to execute the
  computational task and once this is completed the remote MPI task
  transmits the final results to the local. 
\end{itemize}

\paragraph{Timestepping summary}
The computational steps taken to advance the solution from $t=0$ 
to $t=T$ are summarized in Algorithm \ref{doc_dwarf6:alg:sldwarf}.

\begin{algorithm}
\caption{Semi-Lagrangian dwarf timestepping}\label{doc_dwarf6:alg:sldwarf}
\begin{algorithmic}
\State{Read setup parameters (timestep, mesh res, interpolation options etc);}
\State{Setup Atlas mesh;}
\State{Initialize fields;}
\For{step=1,$\ldots$,num\_of\_steps}
\State{Extrapolate in time wind field;}
\State{Set halo fields for wind and tracers;}
\State{Compute the departure points in spherical 3D domain;}
\State{Compute coefficients for interpolation on departure mesh points;}
\State{Interpolate each tracer $\phi_{\chi}$ at the departure point mesh;}
\State{Update fields to new timestep values;}
\EndFor
\end{algorithmic} 
\end{algorithm}

\paragraph{I/O interfaces}
The semi-Lagrangian dwarf requires as inputs a wind field and a tracer field. 
The former is strictly a given input field i.e. it is defined on the entire time integration 
range and it is not changed by the dwarf. The tracer is both an input and output field. 
Its initial value should be specified at the beginning of the integration. This will be 
altered at every timestep as it is advected in the spherical domain.
To summarise, the I/O configuration is as follow.
\begin{itemize}
\item \textbf{Input}: wind field and tracer field.
\item \textbf{Output}: tracer field advected.
\end{itemize}

\subsubsection{Dwarf installation and testing}
In this section we describe how to download and install 
the dwarf along with all its dependencies and we show 
how to run it for a simple test case.

Note that dwarf-D-advection-SemiLagrangian is implemented using ATLAS, 
the ECMWF software framework that supports flexible data-structures for NWP.

\paragraph{Download and installation}
The first step is to download and install the dwarf along 
with all its dependencies. With this purpose, it is possible 
to use the script provided under the ESCAPE software collaboration 
platform:\\
\url{https://git.ecmwf.int/projects/ESCAPE}.

Here you can find a repository called \inlsh{escape}.
You need to download it. There are two options to do this. One option is to use ssh. For this option you need to add an ssh key to your bitbucket account at \url{https://git.ecmwf.int/plugins/servlet/ssh/account/keys}. The link "SSH keys" on this website gives you instructions on how to generate the ssh key and add them to your account. Once this is done you should first create a 
folder named, for instance, ESCAPE, enter into it 
and subsequently download the repository by using the following the steps below:
\begin{lstlisting}[style=BashStyle]
mkdir ESCAPE
cd ESCAPE/
git clone ssh://git@git.ecmwf.int/escape/escape.git
\end{lstlisting}
The other option to download the repo is by using https instead of ssh. Instead of the git command above you then need to use 
\begin{lstlisting}[style=BashStyle]
git clone https://<username>@git.ecmwf.int/scm/escape/escape.git
\end{lstlisting}
where <username> needs to be replace by your bitbucket username.

Once the repository is downloaded into the \inlsh{ESCAPE} folder 
just created, you should find a new folder called \inlsh{escape}. 
The folder contains a sub-folder called \inlsh{bin} that has the 
python/bash script (called \inlsh{escape}) that needs to be 
run for downloading and installing the dwarf and its dependencies. 
To see the various options provided by the script you can type:
\begin{lstlisting}[style=BashStyle]
./escape/bin/escape -h
\end{lstlisting}
To download the dwarf you need to run 
the following command:
\begin{lstlisting}[style=BashStyle]
./escape/bin/escape checkout dwarf-D-advection-SemiLagrangian \ 
--ssh
\end{lstlisting}
To use https you need to replace --ssh with --user <username>. The commands above automatically check out the \inlsh{develop}
version of the dwarf. If you want to download a specific branch 
of this dwarf, you can do so by typing:
\begin{lstlisting}[style=BashStyle]
./escape/bin/escape checkout dwarf-D-advection-SemiLagrangian --ssh \
--version <branch-name>
\end{lstlisting}
Analogous approach can be used for the \inlsh{-\,-user} 
version of the command. You should now have a folder called 
\inlsh{dwarf-D-advection-SemiLagrangian}.

In the above command, you can specify several other optional 
parameters. To see all these options and how to use them you 
can type the following command:
\begin{lstlisting}[style=BashStyle]
./escape checkout -h
\end{lstlisting}

At this stage it is possible to install the dwarf 
and all its dependencies. This can be done in two 
different ways. The first way is to compile and 
install each dependency and the dwarf separately:
\begin{lstlisting}[style=BashStyle]
./escape/bin/escape generate-install dwarf-D-advection-SemiLagrangian
\end{lstlisting}
The command above will generate a script 
called \inlsh{install-dwarf-D-advection-SemiLagrangian} 
that can be run by typing:
\begin{lstlisting}[style=BashStyle]
./install-dwarf-D-advection-SemiLagrangian
\end{lstlisting}
This last step will build and install the dwarf 
along with all its dependencies in the following 
paths:
\begin{lstlisting}[style=BashStyle]
dwarf-D-advection-SemiLagrangian/builds/
dwarf-D-advection-SemiLagrangian/install/
\end{lstlisting}

The second way is to create a bundle that compile 
and install all the dependencies together:
\begin{lstlisting}[style=BashStyle]
./escape/bin/escape generate-bundle dwarf-D-advection-SemiLagrangian
\end{lstlisting}
This command will create an infrastructure to avoid
compiling the single third-party libraries individually
when some modifications are applied locally to one of 
them. To complete the compilation and installation process, 
after having run the above command for the bundle, simply 
follow the instructions on the terminal.

In the commands above that generate the installation 
file, you can specify several other optional parameters. 
To see all these options and how to use them you 
can type the following command:
\begin{lstlisting}[style=BashStyle]
./escape generate-install -h
./escape generate-bundle -h
\end{lstlisting}

\paragraph{Testing}
You should now verify that the dwarf works as expected. For this
purpose, we created a testing framework that allows us to verify that
the main features of the dwarf are working correctly.
In particular, for each sub-dwarf we provide various regression tests
in order to allow the results to be consistent when the underlying
algorithms are modified, and to test additional features or different
hardware. The regression tests can be found 
in the folder test that is located in each sub-dwarf folder. For each
sub-dwarf we also provide scripts running the code on different
architectures, e.g. the Cray 
HPC at ECMWF, that can be found in the folder run-scripts located in each
sub-dwarf folder.

To run this verification, you should run the following 
command:
\begin{lstlisting}[style=BashStyle]
ctest -j<number-of-tasks>
\end{lstlisting}
from inside the \inlsh{builds/dwarf-D-advection-SemiLagrangian}
folder.
\begin{warningbox}
We strongly advise you to verify via ctest that 
the main functionalities of the dwarf are working 
properly any time you apply modifications to the 
code. Updates that do not pass the tests cannot 
be merged. 
In addition, if you add a new feature to the dwarf,
this should be supported by a test if the existing
testing framework is not already able to verify its
functionality.
\end{warningbox}
For instructions on how to run the executables 
see the next section.

\subsubsection{Run the dwarf}

If you want to run the dwarf in your local machine, 
you could do so by using the executable files inside 
\begin{lstlisting}[style=BashStyle] 
install/bin/dwarf-D-advection-SemiLagrangian-prototype1
\end{lstlisting}
In particular, the executables need the specification 
of a configuration file which can be found at
\begin{lstlisting}[style=BashStyle] 
sources/dwarf-D-advection-SemiLagrangian/config-files/
\end{lstlisting}
Configuration file \inlsh{dwarf-D-advection-SemiLagrangian.json} specifies parameters needed
by the dwarf. 
Two test cases are available: (i) a solid body rotation problem
and (ii) the Rossby-Haurwitz problem. To run it you should copy 
the \inlsh{dwarf-D-advection-SemiLagrangian.json} file in the folder
where you  
want to run the simulation (or alternatively specify its path 
on the command line). This file is set to run the solid body rotation
problem using the octahedral mesh O32. This is determined by the value
of the flag \inlsh{init} setting 1 executes solid body rotation
problem while 2 executes Rossby-Haurwitz problem. Other flags of
interest are:
\begin{description}
\item[nlev] number of vertical levels - minimum 4
\item[halo] halo size
\item[iout] output frequency in timesteps (for visualization by gmsh)
\item[dp\_meth] method used to compute departure points [1: Ritchie, 2:
  rotation matrix (default)]
\item[dp\_extrap] time-extrapolation method for wind used in departure
  point calculation [1: standard 2nd order, 2: SETTLS (default)]
\item[interp\_meth] method used to interpolate an advected field at the
  departure point [1: tri-linear, 3: tri-cubic, 4: ECMWF
  quasi-tri-cubic (default)]
\item[lqm] enable quasi-monotone limiter for interp\_meth 3, 4
\item[ndp\_iter] number of iterations in departure point calculation
\item[nsteps] number of timesteps
\item[ntrac] number of advected tracers
\end{description}

There are also a few additional options related to the solid body
rotation problem (setting the winds and the extend of the tracer
field).

The executable can then be run in
\inlsh{dwarf-D-advection-SemiLagrangian/install} directory as follows:
\begin{lstlisting}[style=BashStyle] 
bin/dwarf-D-advection-SemiLagrangian-prototype1 \
--config dwarf-D-advection-SemiLagrangian.json
\end{lstlisting}
where, if the \inlsh{.json} file is not in the current directory, 
you can specify its path after \inlsh{-\,-config}.

If you instead want to run the dwarf on an HPC machine 
available to the ESCAPE partners, you can automatically 
generate the job submission script with the \inlsh{escape} 
file. Python must be available. For convenience,
copy the \inlsh{.json} input file to a file with name
\inlsh{input.json} in the escape directory and execute:
\begin{lstlisting}[style=BashStyle]
./escape/bin/escape generate-run -c \
dwarf-D-advection-SemiLagrangian/install/dwarf-D-advection-SemiLagrangian/bin
/dwarf-D-advection-SemiLagrangian-prototype1 \
\end{lstlisting}
This will generate the submission script 
for the given HPC machine you are targeting without submitting 
the actual job. The command above will in fact simply generate 
an \inlsh{escape.job} file in the current folder. This can
successively be submitted via \inlsh{qsub} on the HPC machine 
you want to run the simulation on.

In the above command you can specify several other optional 
parameters, such as wall-time, number of tasks, number of 
threads, etc. To see all these options and how to set them 
up you can type the following command:
\begin{lstlisting}[style=BashStyle]
./escape/bin/escape generate-run -h
\end{lstlisting}

\subsubsection{Integration}

This dwarf explores the solution of the advection problem in
a spherical domain arising in the context of prognostic
equations of NWP. However it is not sufficient on
its own to solve these equations as they include other
forcing terms beyond the advection terms. Given that some of the extra
terms give rise to fast moving waves, typically, a semi-implicit
method of integration is required to allow maintaining the stability
advantage of the SL method of using long timesteps. In particular,
this dwarf can be integrated in a global Weather \& Climate model
involving a semi-implicit time-stepping scheme and a mesh-based 
spatial discretization. A 3D elliptic equation solver would normally
be required for a gridpoint model to solve efficiently the derived
Helmholtz equation which is part of a standard semi-implicit
method. For a spectral transform model this is not needed provided that
a constant coefficient formulation is used for the Helmholtz problem.